\documentclass[10pt,conference]{IEEEtran}
\IEEEoverridecommandlockouts
\usepackage{cite}
\usepackage{amsmath,amssymb,amsfonts}
\usepackage{algorithmic}
\usepackage{graphicx}
\usepackage{textcomp}
\usepackage{xcolor}
\usepackage{subfigure}
\usepackage{xspace}
\usepackage{mathtools}
\usepackage{enumitem}
\usepackage{multirow}
\usepackage{adjustbox}
\usepackage{tikz}
\usepackage{siunitx}
\usepackage{caption}
\usepackage{verbatim}

\def\BibTeX{{\rm B\kern-.05em{\sc i\kern-.025em b}\kern-.08em
    T\kern-.1667em\lower.7ex\hbox{E}\kern-.125emX}}

\newcommand{\tool}{KNOD\xspace}
\newcommand{\tooldecoder}{KNOD$_{\text{-decoder}}$\xspace}
\newcommand{\toolinf}{KNOD$_{\text{-distInf}}$\xspace}
\newcommand{\tooltrain}{KNOD$_{\text{-distTrain}}$\xspace}

\newcommand{\decoder}{three-stage tree decoder\xspace}
\newcommand{\domain}{domain-rule distillation\xspace}
\newcommand{\Domain}{Domain-Rule Distillation\xspace}
\newcommand{\rules}{logic rules\xspace} %type rules
\newcommand{\drules}{rules\xspace} %distillation rules\xspace} 
\newcommand{\astgraph}{ASG}
\newcommand{\asg}{ASG}

\newcommand{\coconut}{CoCoNuT\xspace}
\newcommand{\cure}{CURE\xspace}
\newcommand{\codit}{CODIT\xspace}
\newcommand{\recoder}{Recoder\xspace}
\newcommand{\dlfix}{DLFix\xspace}
\newcommand{\sequencer}{SequenceR\xspace}
\newcommand{\rewardrepair}{RewardRepair\xspace}

\newcommand{\tbar}{TBar\xspace}

\newcommand{\defects}{Defects4J\xspace}
\newcommand{\defectsold}{Defects4J v1.2\xspace}
\newcommand{\defectsnew}{Defects4J v2.0\xspace}
\newcommand{\quixbugs}{QuixBugs\xspace}

\newcommand{\javalang}{\texttt{javalang}\xspace}
\newcommand{\javaparser}{\texttt{JavaParser}\xspace}

\newcommand{\todoc}[2]{{\textcolor{#1}{\textbf{#2}}}}

\newcommand{\todogreen}[1]{\todoc{green}{\textbf{[[#1]]}}}
\newcommand{\todoblue}[1]{\todoc{blue}{\textbf{[[#1]]}}}

\newcommand{\todobrown}[1]{\todoc{brown}{\textbf{[[#1]]}}}

\newcommand{\lin}[1]{\todoblue{Lin: #1}}

\newcommand{\nan}[1]{\todobrown{Nan: #1}}

\newcommand{\thibaud}[1]{\todogreen{Thibaud: #1}}

\renewcommand{\todoc}[2]{\relax}
\newcommand{\rev}[1]{{#1}}

\newcommand{\nsty}[1]{\textit{#1}} % node style
\newcommand{\esty}[1]{\textit{#1}} % node style 
\newcommand{\code}[1]{\texttt{\small #1}} % code style 

\newcommand*\circled[1]{\tikz[baseline=(char.base)]{
            \node[shape=circle,draw,inner sep=1pt] (char) {#1};}}

\newcommand{\distance}{8pt}
\begin{document}

\title{\tool: Domain \underline{Kno}wledge \underline{D}istilled Tree Decoder for Automated Program Repair}

\author{
\IEEEauthorblockN{Nan Jiang}
\IEEEauthorblockA{
\textit{Purdue University}\\
West Lafayette, USA \\
jiang719@purdue.edu} \\
\IEEEauthorblockN{Lin Tan}
\IEEEauthorblockA{
\textit{Purdue University}\\
West Lafayette, USA \\
lintan@purdue.edu}
\and
\IEEEauthorblockN{Thibaud Lutellier}
\IEEEauthorblockA{
\textit{University of Alberta$^1$}\thanks{$^1$This work is done when Thibaud was at University of Waterloo.}\\
Alberta, Canada \\
lutellie@ualberta.ca} \\
\IEEEauthorblockN{Dan Goldwasser}
\IEEEauthorblockA{
\textit{Purdue University}\\
West Lafayette, USA \\
dgoldwas@purdue.edu}
\and
\IEEEauthorblockN{Yiling Lou}
\IEEEauthorblockA{
\textit{Fudan University$^2$}\thanks{$^2$This work is done when Yiling was at Purdue University.}\\
Shanghai, China \\
yilinglou@fudan.edu.cn} \\
\IEEEauthorblockN{Xiangyu Zhang}
\IEEEauthorblockA{
\textit{Purdue University}\\
West Lafayette, USA \\
xyzhang@cs.purdue.edu}
}

\maketitle

\begin{abstract}
Automated Program Repair (APR) improves software reliability  by generating patches for a buggy program automatically. Recent APR techniques leverage deep learning (DL) to build models to learn to generate patches from existing patches and code corpora. 
While promising, DL-based APR techniques suffer from the abundant syntactically or semantically incorrect patches in the patch space. These patches often disobey the syntactic and semantic \emph{domain knowledge} of source code and thus cannot be the correct patches to fix a bug. 
% Therefore, incorporating the \emph{domain knowledge} into DL-based APR techniques could help avoid generating such invalid patches to improve the effectiveness and efficiency of patch generation. 

We propose a DL-based APR approach \emph{\tool{}}, which incorporates domain knowledge to guide patch generation in a \emph{direct and comprehensive} way.
\tool{} has two major novelties, including (1) a novel \emph{\decoder{}},  which \emph{directly} generates Abstract Syntax Trees %(ASTs) 
of patched code according to the inherent tree structure,
%in three stages with \emph{three decoders}
and (2) a novel \emph{domain-rule distillation}, which leverages syntactic and semantic rules and teacher-student distributions to explicitly inject the domain knowledge into the decoding procedure during \emph{both the training and inference phases}. 

We evaluate \tool{} on three widely-used benchmarks. 
%\defectsold{}, \defectsnew{}, and \quixbugs{}.
\tool{} fixes 72 bugs on the \defectsold{}, 25 bugs on the \quixbugs{}, and 50 bugs on the additional \defectsnew{} benchmarks, outperforming all existing APR tools.
%\tool also fixes the second most, 42, on the additional \defectsnew{} bugs, which shows \tool's generalizability. 
%In addition, our ablation study confirms the contribution of both novelties in our domain-knowledge-distilled tree-decoder architecture.
\end{abstract}


\begin{tikzpicture}[remember picture,overlay]
    \node[anchor=south,yshift=765pt] at (current page.south) {\parbox{\dimexpr 1\textwidth-\fboxsep-\fboxrule\relax}{\centering 2023 IEEE/ACM 45rd International Conference on Software Engineering (ICSE)}};
\end{tikzpicture}
\newcommand\copyrightnotice[1]{
    \begin{tikzpicture}[remember picture,overlay]
    \node[anchor=south,yshift=40pt] at (current page.south) {\parbox{\dimexpr 1\textwidth-\fboxsep-\fboxrule\relax}{\footnotesize #1}};
    \end{tikzpicture}
}
\copyrightnotice{
xxxxx~\copyright~2023~IEEE. \\ 
DOI xxxxx
}

\section{Introduction}
\noindent Since developers spend nearly half of their time fixing bugs (49\%$\pm$39\%)~\cite{debug}, support to help developers fix bugs  is in high demand. Automated program repair~\cite{DBLP:journals/cacm/GouesPR19,DBLP:journals/csur/Monperrus18,livingapr} exactly provides such support, which %is a crucial research domain that aims at 
generates patches for buggy programs with little manual effort to improve software reliability and reduce software development costs. %Researchers have proposed various strategies to guide patch generation, including heuristics-based~\cite{DBLP:journals/tse/YuanB20/arja,DBLP:conf/icse/WenCWHC18,DBLP:conf/kbse/SahaLYP17/elixir}, constraint-based~\cite{DBLP:conf/icse/XiongWYZH0017/acs, DBLP:journals/tse/XuanMDCMDBM17/nopol,DBLP:conf/ssbse/MartinezM18/cardumen}, and learning-based approaches~\cite{rewardrepair, chen2018sequencer, lutellier2020coconut,jiang2021cure,hopity, zhu2021recoder}. 
\lin{I removed this sentence from intro. please make sure relevant papers are cited in the paper (they probably are already cited in related, but pls double check.}
With the rapid development of deep learning, recent learning-based APR techniques~\cite{rewardrepair, chen2018sequencer, lutellier2020coconut,jiang2021cure,hopity, zhu2021recoder,transfer-vul,selfapr} leverage advanced DL techniques to generate patches by learning from existing code corpora. %During the training phase, these techniques build a model from existing patches and software repositories, and during the inference phase, these techniques use the model to generate candidate patches.
DL-based APR often formulates APR as the translation from the given buggy code to the correct one and adopts neural machine translation (NMT) techniques. They typically follow an encoder-decoder architecture, where the encoder first embeds the buggy code, while the decoder generates the patched code iteratively.  
The generated patches are then validated against test cases.

One major challenge of DL-based APR is that invalid (i.e., 
syntactically or semantically incorrect) patches dominate in the patch space~\cite{lutellier2020coconut, jiang2021cure,zhu2021recoder,rewardrepair}, which hurts the effectiveness and efficiency of patch generation.  
This challenge is exacerbated because traditional DL encoders and decoders are designed and built for input such as images and text instead of source code. 
Different from images and text, source code is a formal language with its own syntax and semantics. Thus, patches disobeying such syntactic and semantic \emph{domain knowledge} cannot be correct patches to fix a bug. 

Ideally, given source code and patches as training data, one expects traditional DL models to learn code syntax and semantics well. However, in practice, such models generate a large portion of uncompilable and incorrect patches~\cite{lutellier2020coconut,jiang2021cure,zhu2021recoder, chen2018sequencer}, wasting a daunting amount of computing power and, in many cases preventing correct patches from being generated with bounded resources. Thus, it is crucial to enforce syntactic and semantic rules directly on DL models. In addition, such rules must be enforced during the training phase, as opposed to during inference only (e.g., filtering out generated patches that cannot be parsed or type-checked), because enforcing syntax and semantics during inference still causes a large portion (as high as 91\%) of uncompilable and incorrect patches~\cite{jiang2021cure, zhu2021recoder}.

To leverage domain knowledge to guide DL-based APR in a \emph{direct and comprehensive} way, we propose a novel DL-based APR approach---\textbf{\tool{}}, consisting of a novel three-stage decoder with domain-rule distillation. \tool{} has two major novelties. First, different from previous work that generates sequences or production rules~\cite{lutellier2020coconut,jiang2021cure, zhu2021recoder, chen2018sequencer}, our \textbf{three-stage decoder} \emph{directly} generates  ASTs of patched code from root to children according to the AST tree structure with \emph{three decoders}: a parent decoder, an edge decoder, and a node decoder. Such a three-stage design enforces the model to naturally capture the tree structure in the ASTs, helping the model learn AST syntax and semantics. 

Second, the decoder incorporates a \textbf{domain-rule distillation} component to explicitly inject the domain knowledge into the decoding procedure. Specifically, the domain-rule distillation component first represents syntax and semantics as rules expressed in first-order logic (FOL). It then uses these \rules{} to refine the teacher-student probability distributions to guide our model to learn to follow these syntactic and semantic rules.
%, i.e., by modifying model weights. 
Different from existing work, our domain-rule distillation  uses \rules{} to \emph{explicitly} modify  the optimization function in the decoding procedure in \emph{both the training and inference phases}, which thus should have a stronger capability of utilizing domain knowledge. 

In summary, this paper makes the following contributions.
\begin{itemize}
    \item \textbf {A \decoder} to directly generate Abstract Syntax Trees (ASTs) in three stages to capture tree structures, syntax, and semantics,  
    \item \textbf{A \domain during training and inference} to explicitly modify the optimization function %during both the training and inference phase 
    to guide the \decoder to follow code syntax and semantics, 
    \item \textbf{An APR technique \tool{}} based on the proposed domain-knowledge-distilled tree-decoder architecture, 
    %including a novel \emph{\decoder{}} with \emph{\domain{}}, and  
    \item \textbf{An  evaluation} of \tool on
    three widely-used benchmarks, \defectsold, \defectsnew~\cite{defects4j}, and \quixbugs~\cite{quixbugs}. 
    %\tool{} is consistently effective on all three benchmarks. Specifically, 
    \tool{} outperforms all existing non-DL and DL-based approaches by fixing 72 bugs on \defectsold{}, fixing 8 and 19 more bugs than the most effective DL-based and non-DL-based APR techniques, respectively. \tool also fixes the most bugs, 25 and 50, on the \quixbugs and \defectsnew benchmarks, which shows \tool's generalizability. 
    % Our ablation study shows that each component of incorporating domain knowledge indeed contributes to the effectiveness of \tool{}. 
\end{itemize}

%\smallskip
% \noindent\textbf{Availability:}  Our replication package is available~\cite{share}. 
% \todo{update the shared repo}

\section{Approach}
\noindent This section presents our proposed approach, \tool. 
Section~\ref{sec:overview} gives an overview of \tool. 
Section~\ref{sec:data representation} shows how \tool{} represents code;
Sections~\ref{sec:encoder} to~\ref{sec:domain knowledge fusion} describes the DL models;
Section~\ref{sec:trainandinter} presents training and inference; Section~\ref{sec:patch generation} describes patches' generation and validation. 

\subsection{Overview}
\label{sec:overview}

\begin{figure*}
    \centering
    \includegraphics[width=0.9\linewidth]{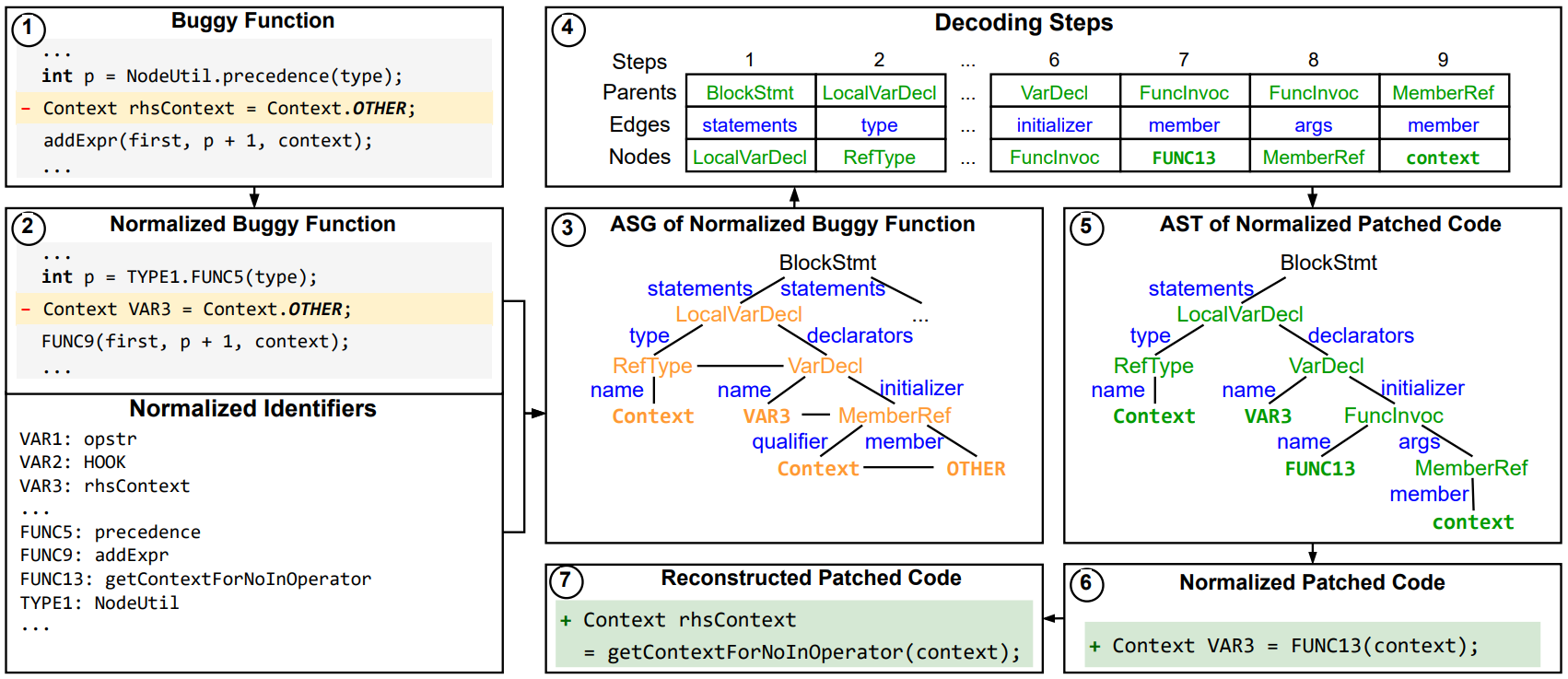}
    \caption{An example of \tool fixing bug Closure-123 in \defects. Stmt stands for Statement, Var is Variable, Decl is Declaration,  Ref is Reference, Func is Function, and Invoc is Invocation.
 }
    \label{fig:ast example}
\vspace{-1em}
\end{figure*}

\noindent \tool{} consists of two phases: the training phase and the inference phase. During the training phase, \tool{} takes buggy code and its patches  as input and trains a model (an encoder and a decoder) to learn how to generate patches to fix bugs automatically. During the inference phase, \tool{} takes an unseen buggy project, including the location of buggy lines as input, which are standard input that existing APR techniques take~\cite{jiang2021cure,lutellier2020coconut,rewardrepair,chen2018sequencer, zhu2021recoder}. \tool's trained  decoder generates ASTs, which are then reconstructed into patched code.\thibaud{Where is the "construction component" in the rest of the section?}\nan{fixed} \tool{} then automatically validates these code patches with test cases to generate candidate patches for developers to review. 

Figure~\ref{fig:ast example} illustrates how \tool{} fixes the Closure-123 bug in the widely-used bug benchmark \defects{}. The Closure-123 bug is uniquely fixed by \tool{}. Given the buggy function and the bug location (step \circled{1}, where the bug is in a yellow background), \tool{} normalizes the buggy function by replacing uncommon identifiers with normalized textual representations (step \circled{2}, Section~\ref{sec:data representation}) and then builds the Abstract Syntax Graph (ASG) of the normalized buggy function (step \circled{3}, Section~\ref{sec:data representation}). % Section~\ref{sec:data representation} ``ASTs and Abstract Syntax Graphs")
The ASG is the input to \tool{}'s APR model, which is trained to decode the AST of a normalized patch (steps \circled{4} and \circled{5}, Sections~\ref{sec:encoder} $\sim$ \ref{sec:trainandinter}). 
\tool transforms the AST into a patch in its normal form (step \circled{6}, Section~\ref{sec:patch generation}), which is eventually instantiated into a real code patch by replacing the abstract tokens with the concrete identifiers  (step \circled{7}, Section~\ref{sec:patch generation}).

\begin{figure*}[t]
  \centering
  \includegraphics[width=0.9\linewidth]{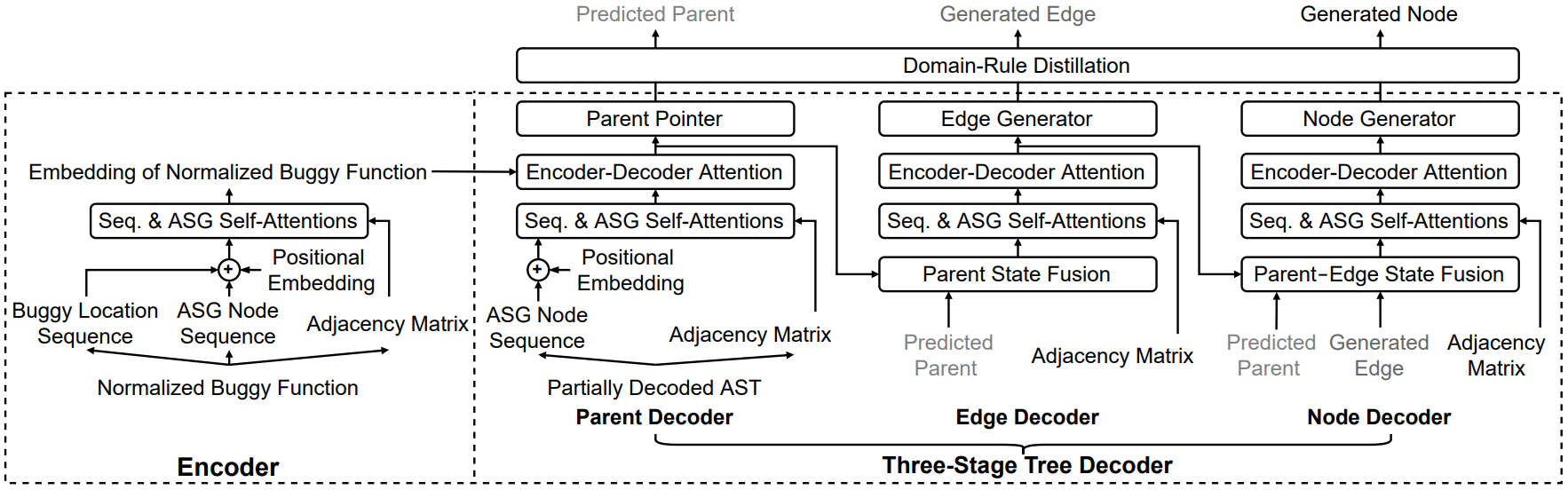}
  \caption{Architecture of \tool's  model, including a graph-transformer encoder, a \decoder (parent, edge, and node decoder), and a \domain module.}
  \label{fig:architecture}
\end{figure*}

Figure~\ref{fig:architecture} presents the  overview of \tool's 
%encoder-decoder architecture, which  consists of three main components: (i) an \emph{encoder} for  buggy functions (Section~\ref{sec:encoder}), (ii) a novel \emph{\decoder} for ASTs of patched code (Section~\ref{sec:decoder}), and (iii) a \emph{\domain} that learns code syntactic and semantic information to improve patch correctness (Section~\ref{sec:domain knowledge fusion}). 
\textbf{domain-knowledge-distilled tree-decoder architecture} which has two main novelties. 

\smallskip
\noindent\textbf{Novelty 1: \decoder{}.} First, different from existing APR work that generates sequences or production rules, our \emph{\decoder{}} generates bug fixes in an AST format directly  (Section~\ref{sec:decoder}), which is then automatically converted to source code  (Section~\ref{sec:patch generation}). 
Specifically, our \decoder{} includes three decoders:  (i) a parent decoder, which  selects the parent node among generated nodes to work on at each step, (ii) an edge decoder, which generates an edge  for the parent node that was selected by the parent decoder, and generates the label of the edge (differ from standard ASTs, our ASTs have labels for edges to be leveraged by models as detailed in Section~\ref{sec:data representation} ``ASTs and Abstract Syntax Graphs"), 
and (iii) a node decoder, which generates a new node connected to the edge generated by the edge decoder. 
\emph{Different from existing APR decoders that generate patches as token sequences~\cite{jiang2021cure, lutellier2020coconut,chen2018sequencer}, our decoder generates patches in ASTs with explicit structure. As such, our model is forced to learn explicit code structure. Compared to those generating patches as grammar production rules, our decoder emits AST edges with explicit labels, distinguishing various edge labels. In addition, tree generation is incremental, with one edge and one node emitted at a time. Such a design enables fine-grained control over the quality of generated trees (e.g., enforcing syntactic and semantic validity during generation).}
 
Using Figure~\ref{fig:ast example}\circled{4} as an example, \tool's decoder starts from the root node \nsty{BlockStmt}, which is always the same as the root node of the buggy AST (\circled{3}). In step 1, our decoder selects \nsty{BlockStmt} as the parent node for this step, generates an edge with label \esty{statements} for this parent, and then generates the next node \nsty{LocalVarDecl}. 
In step 2, our decoder selects node   \nsty{LocalVarDecl} as the parent,  generates an edge of label \esty{type}, and the next node \nsty{RefType}. The resulting AST is in Figure~\ref{fig:ast example}\circled{5}.
Section~\ref{sec:rq2} shows that with a \decoder, \tool fixes more bugs.

\smallskip
\noindent\textbf{Novelty 2: \domain during training and inference.} 
Second, a straightforward design is to first use \tool's decoder to generate many candidate patch ASTs and then use the parser, type checker, and test suite to rule out the invalid ones, similar to existing APR techniques~\cite{jiang2021cure,zhu2021recoder,lutellier2020coconut,chen2018sequencer}. However, without training the decoder to generate syntactically and semantically valid patches, most of the generated ASTs are invalid, which incurs substantial validation overhead and prevents correct fixes from being  generated with bounded efforts.

Therefore, a critical design choice of \tool{} is \emph{\domain}, which  enforces \tool's decoder to generate valid patches during training in addition to the inference phase. Specifically, during training, we universally encode the grammar and type-checking rules as first-order logic formulas. 
We create a teacher distribution using these \rules to modify the student distribution, which is the distribution from our encoder-decoder model. \emph{We create a loss between the student distribution (model distribution) and the teacher distribution (model distribution with syntactic and semantic rules), and add this loss to penalize ASTs that violate the grammatical or type rules.} This loss is added to the traditional loss used by APR techniques, which is between the student distribution (model distribution) and the ground-truth labels (correct developer patches in training data). 
%Our \domain{} uses the \rules{} to update the teacher-student distributions at each iteration of node/edge generation to force the encoder-decoder model to learn the \rules{}. 
\emph{The teacher-student loss enables domain knowledge transfer to the models' weights at each iteration}~\cite{firstlogic}. In summary, our \domain{} enables our APR models to learn from both the training data and the \rules{}. During inference, the model automatically distills the invalid generations and emits the likely valid ones. 

For the Closure-123 bug in Figure~\ref{fig:ast example}, an example semantic rule is that
when the model  generates a member (e.g., \code{context}) as an argument of a function invocation (e.g., \code{FUNC13(context)}), only nodes whose types are compatible with that function's argument type should be generated.  Figure~\ref{fig:ast example}\circled{2} shows that \code{FUNC13} was normalized from \code{getContextForNoInOperator}, whose argument type  is \code{Context}. Thus, only identifiers whose types are compatible with type \code{Context} are valid. Our \domain sets the  probabilities of all  identifiers with incompatible types to 0. As a result, the decoder can generate the correct node \code{context} instead of \code{p} (of type \code{int}), which has the highest probability without \domain. 
% A vanilla decoder generates \code{p}, which fails to fix this bug. 

An example syntactic rule is that node \nsty{MemberRef} must have only one edge of label \esty{member} (for member reference, there must be one and only one member). 
Since our \domain modifies the distributions during training (similar to the example above to set the probabilities of invalid edge/node labels to 0), which teaches the model (by modifying model weights) to give the edge of label \esty{member} the highest probability for the parent node \nsty{MemberRef} during inference, \tool generates a valid AST. 
%Without \domain, a vanilla decoder generates other  edge labels for \nsty{MemberRef}, and fails to fix this bug. 
Section~\ref{sec:rq2} shows that with \domain{}, \tool{} fixes more bugs.

\subsection{Data Preprocessing and Extraction}
\label{sec:data representation}

\noindent \tool normalizes a given buggy function and its patch and parses them into Abstract Syntax Graphs, which are directional graphs with edges added to connect sibling nodes in an AST.  Such additional edges enable closer distance between sibling nodes to help models learn syntaxes and semantics effectively. While the edges between siblings nodes are useful for the encoder to facilitate learning, they are not part of source code and do not need to be generated by our decoder,  because it is trivial to add such edges to a generated AST. Thus, we only need an AST decoder. 

\smallskip\noindent{}\textbf{Code Normalization.} 
Due to the potentially unlimited number of unique identifiers in source code, NMT-based APR models~\cite{chen2018sequencer,jiang2021cure,lutellier2020coconut}
usually suffer from the large vocabulary size issue and the out-of-vocabulary problem. Therefore, to address these issues, \tool{} first performs code normalization by applying src2abs~\cite{tufano2018src2abs, tufano2019src2abs} to transform identifiers (e.g., data types, function names, variable names, and literal values) to  normalized textual representation. For example, the buggy function for bug Closure-123 (Figure~\ref{fig:ast example}\circled{1}) is transformed to normalized code (Figure~\ref{fig:ast example}\circled{2}), where the mappings from normalized identifiers to concrete identifiers are kept for reconstruction later. In addition, we denote identifiers that appear only in the patch but not in the buggy function as unknown identifiers, which are normalized to special placeholders (e.g., ``\code{TYPE-UNK}'', ``\code{FUNC-UNK}'', and ``\code{VAR-UNK}'').

\smallskip
\noindent{}\textbf{ASTs and Abstract Syntax Graphs.} 
After parsing the normalized code of the buggy function and the patched line into ASTs, \tool{} then represents these ASTs as \emph{Abstract Syntax Graphs}. \astgraph{} is a directional graph whose vertices are the AST nodes, while edges are (i) the original edges between AST nodes, and (ii) the new edges between each AST node and its siblings, which make them graphs instead of trees.
For example, Figure~\ref{fig:ast example}\circled{3} illustrates the \astgraph{} of the normalized buggy function of the Closure-123 bug. 
In classic ASTs, edges are introduced between a nonterminal symbol on the left-hand side of a %BNF 
production rule (i.e., the parent) to  the symbols, terminal or nonterminal, on the right-hand side of the rule (i.e., the children). 
While these edges are often not labeled, they have different semantics, which can be leveraged in model training. Thus, we label these edges explicitly. For example, in 
Figure~\ref{fig:ast example}\circled{3}, node \nsty{LocalVarDecl} has two child nodes---node \nsty{RefType} denoting a reference type (i.e., \code{Context}), and node \nsty{VarDecl}  denoting a variable declaration. 
The two edges $\mathit{LocalVarDecl}\rightarrow\mathit{RefType}$ and $\mathit{LocalVarDecl}\rightarrow\mathit{VarDecl}$ have different meanings. Therefore, we introduce explicit edge labels, such as \esty{type} and \esty{declarators} for the aforementioned two edges. These labels provide different granularities of information compared to  node labels. For instance, a {\it type} edge may lead to various child nodes such as \nsty{RefType} and \nsty{BasicType}. 
We use the additional information carried by edge labels to improve the model. For example, given the parent \nsty{LocalVarDecl} node, without the label \esty{type}, a decoder could generate invalid nodes such as \code{FUNC5}. In contrast, with the label \esty{type}, a decoder could learn to focus on generating type nodes such as \nsty{RefType}. We use the \javalang parser to directly extract and create these edge and node labels from code. Our design is unique compared to the state-of-the-art~\cite{zhu2021recoder,hopity}, which do not use explicit edge labels. 

\subsection{Encoder} 
\label{sec:encoder}

\noindent During both training and inference, our encoder takes ASGs of normalized buggy functions as input and output embeddings of these functions to be used by our decoder. We chose graph-transformer~\cite{Yun2019graphtransformer,Vijay2020graph-transfrmer} as the architecture for our encoder, given its superior scalability and learning capability on graph-structured data.  
To improve the encoder's effectiveness, we apply \emph{sequence self-attention} and \emph{ASG self-attention} to the encoder to highlight dependencies between every node pair and every adjacent ASG node pair.

Following existing work ~\cite{Vijay2020graph-transfrmer,Yun2019graphtransformer,treedec-markup,zhu2021recoder}, we use two vectors/matrices (``ASG Node Sequence" and ``Adjacency Matrix" in Figure~\ref{fig:architecture}) to represent each ASG, which are the input formats that a deep learning encoder can take.  
The ASG Node Sequence, notated by $\{n_i\}=(n_1, n_2,\ldots, n_L)$, is a sequence of node names following a pre-order traverse of \asg{} (e.g., [\nsty{BlockStmt}, \nsty{LocalVarDecl}, ...]). The adjacency matrix $A=\{a_{ij}\}$ keeps the edge information where $a_{ij}=0$ means there is no edge between node $n_i$ and $n_j$, while $a_{ij}=k$ means nodes $n_i$ and $n_j$ are linked by an edge whose label is $k$ (e.g., \esty{statements}). 
The remaining input to the encoder is ``Buggy Location Sequence'' (Figure~\ref{fig:architecture}). To let the encoder know which nodes in the ASG node sequence belong to the buggy line, \tool locates the \asg{} nodes belonging to the buggy line, and generates them as a ``Buggy Location Sequence"  $\{t_i\}=(t_1, t_2, \ldots, t_L)$ that follows the same order as ${n_i}$, where $t_i=1$ means node $n_i$ belongs to the buggy lines and  $t_i=2$ means node $n_i$ belongs to non-buggy lines.
%(e.g., the yellow nodes in Figure~\ref{fig:ast example}\circled{3} belong to the buggy line and thus have the buggy tag $t_i=1$, and the rest nodes such as \nsty{BlockStmt} have buggy tag $t_i=2$).

The \asg{} node sequence, adjacency matrix, and buggy location sequence are converted to vectorized embeddings for later computation. Respectively, $\{\boldsymbol{n_i}\}$ is the sequence of \asg{} node embedding, $\boldsymbol{A}=\{\boldsymbol{a_{ij}}\}$ is the embedding of edges in the adjacency matrix, and $\{\boldsymbol{t_i}\}$ is the sequence of buggy location embedding (in this paper, we use light letters for names, labels, etc., use \textbf{bolded} letters for vectors, embeddings, etc., and use $\{ \}$ for sequence). To let the model learn positional and order information, a positional embedding~\cite{transformer,position} $\{\boldsymbol{p_i}\}$ is used to encode the absolute index of each node (1, 2, etc.) in the \asg{} node sequence. All the embeddings are learned by the encoder, and each node $n_i$ is represented by a vector $\boldsymbol{e_i}=\boldsymbol{n_i}+\boldsymbol{t_i}+\boldsymbol{p_i}$.

The encoded vectors $\{\boldsymbol{e_i}\}$ are then fed to a stack of encoder blocks (the number of encoder blocks is configurable), and each encoder block contains a \emph{sequence self-attention module} and an \emph{ASG self-attention module}.

\smallskip
\noindent \textbf{Attention in Encoder.}
The self-attention module follows the standard architecture in existing architectures~\cite{transformer,Yun2019graphtransformer,Vijay2020graph-transfrmer}. The sequence self-attention captures the dependencies between embeddings of every node pair in the \asg{} node sequence, which is widely used in sequential transformer-based neural networks~\cite{transformer}. The ASG self-attention explicitly highlights the dependencies between adjacent nodes. Different from traditional tree self-attention, which considers node features only, our ASG self-attention considers both node and edge features. The attention-based hidden states later are fed to a normalization layer~\cite{layernorm,transformer} and a feed-forward layer~\cite{transformer} to get the final encoder output $\{\boldsymbol{h^{e}_i}\}$, which are the  hidden states for each node in \asg{} for the input buggy function.

\subsection{Novelty 1: Three-Stage Tree Decoder}
\label{sec:decoder}

\noindent Decoding ASTs (more generally, decoding trees) for modern languages such as Java is challenging and less studied. Since modern programming languages such as Java are not context-free, a straightforward approach to decoding trees defined by a context-free grammar that constructs a tree from its root to leaves based on production rules does not work well for Java~\cite{treedec-nmt}. 
% A feasible approach is to decode ASTs directly from embedding, like some traditional tree decoders~\cite{treedec-markup} in the DL domain, instead of constructing them following production rules. 
Besides, ASTs are domain-specific trees with labels on edges, e.g., \esty{declarators} and \esty{initializer} in Figure~\ref{fig:ast example}. Thus, a decoding method needs to decode the edge labels properly. Thus, we  propose a novel \decoder that decodes AST nodes and edges iteratively. 
Specifically, our \decoder includes three sub-decoders: 
(i) a parent decoder, which  selects the parent node among generated nodes to work on at each step, 
(ii) an edge decoder, which generates an edge  for the parent node selected by the parent decoder, and generates the label of the edge, and 
(iii) a node decoder, which generates a new node connected to the edge generated by the edge decoder.

%\emph{Different from existing APR decoders~\cite{jiang2021cure,zhu2021recoder,lutellier2020coconut,chen2018sequencer} that generate sequences or production rules, our decoder generates AST trees directly} (detailed comparison with related work in Section~\ref{sec:related}). Below we introduce each decoder in detail. 

\subsubsection{Parent Decoder}
\label{sec:parent decoder}
The parent decoder, as the first part of the \decoder, takes three inputs, including (i) the encoder's output $\{\boldsymbol{h^e_i}\}$, (ii) the node vectors $\boldsymbol{e^d_i}$ (i.e., $\boldsymbol{e^d_i} = \boldsymbol{n^d_i} + \boldsymbol{p^d_i}$, the sum of the embedding of the node sequence $\{\boldsymbol{n^d_i}\}$ and the positional embedding $\boldsymbol{p^d_i}$), and (iii) the adjacency matrix $\{\boldsymbol{a^d_{ij}}\}$ of the partially generated AST (i.e., the partial AST that the decoder has generated from the beginning to the current decoding iteration). 
As shown in Figure~\ref{fig:architecture}, the parent decoder contains a sequence self-attention, an ASG self-attention, encoder-decoder attention, and a parent pointer.

\smallskip\noindent\textbf{Attention in Parent Decoder.}
The parent decoder leverages the same sequence self-attention and AST self-attention as the ones used in the encoder. However, the attention in the decoder is computed among the nodes in the AST of the patch code.
%(i.e., $\boldsymbol{Q}$, $\boldsymbol{K}$ and $\boldsymbol{V}$ are set to $\{\boldsymbol{e^d_i}\}$)
%As for the encoder-decoder attention, $\boldsymbol{Q}$ is set to $\{\boldsymbol{e^d_i}\}$, while $\boldsymbol{K}$ and $\boldsymbol{V}$ are set to $\{\boldsymbol{h^e_i}\}$.
Besides, the encoder-decoder attention computes the attention weights from the nodes in the patch AST to the nodes in the input buggy AST, which could capture the dependencies between the patch code and the buggy code. The outputs are then normalized by a normalization layer and projected by a feed-forward layer. In this way, for each node in the patch AST, we get its hidden states $\{\boldsymbol{h^{d(p)}_i}\}$ output by the parent decoder (superscript $(p)$ means the hidden states' output by the \emph{parent} decoder).

\smallskip\noindent\textbf{Parent Pointer.}
The parent pointer component leverages the hidden states' output by the last attention module to locate the parent node, for which a new child node would be generated. In particular, the parent point~\cite{pointer} selects an existing node in the partial patch AST as the parent node.
$\text{P}^{(p)}_i(j)$, the probability distribution of the $j$-th nodes being the parent at the $i$-th decoding iteration is
\begin{equation}
\small{
    \text{P}^{(p)}_i(j) = \: \text{softmax}\big(\dfrac{W_q\boldsymbol{h^{d(p)}_i} \cdot W_k\boldsymbol{h^{d(p)}_j}}
    {\sqrt{d}}\big)
}
\end{equation}
where $W_q$ and $W_k$ are trainable weights, and $d$ is the dimension of hidden states.
Intuitively, the parent node with the index $p^d_i = \text{argmax}_j\text{P}^{(p)}_i(j)$ at the decoding iteration $i$, should be the node with the highest attention weight to the $i$-th node in the AST of normalized patched code. 

\subsubsection{Edge Decoder}
\label{sec:edge decoder}
After locating the parent node $\{p^d_i\}$ in the current AST,  the edge decoder component  predicts the label of the new edge that connects the parent node and its child node to be generated. The edge decoder is supposed to work based on the prediction results of the previous parent decoder, since different parent nodes should have different predictions of edge labels. For example, if the parent node is an \nsty{IfStmt},  the edge decoder should predict an edge label of \esty{condition}; if the parent node is a \nsty{FuncInvoc},  the edge decoder should predict an edge label of  \esty{member}. Therefore, the edge decoder takes two inputs, including (i) the parent decoder's output hidden states $\{\boldsymbol{{h^{d(p)}_i}}\}$, and (ii) the predicted parent index $\{p^d_i\}$. 
As shown in Figure~\ref{fig:architecture}, the edge decoder contains a \emph{parent states fusion layer}, followed by three attention modules,
and an edge generation component. The attention modules (i.e., sequence self-attention, ASG self-attention, and encoder-decoder attention) in the edge decoder are similar to those in the parent decoder. Therefore, we focus on describing the novel parent state fusion component and the edge generator as follows.

\smallskip\noindent\textbf{Parent State Fusion.}
\tool{} incorporates a parent state fusion layer to combine the hidden states of each node $\{\boldsymbol{h^{d(p)}_i}\}$ and the hidden states of its predicted parent node $p^d_i$:
$ \boldsymbol{h^{d(e)}_i} = W_f\big(\boldsymbol{h^{d(p)}_i} + \boldsymbol{h^{d(p)}_{p^d_i}}\big)$
, where $W_f$ is a trainable parameter for parent states fusion.

\smallskip\noindent\textbf{Edge Generator.}
Given the hidden states $\{\boldsymbol{h^{d(e)}_i}\}$ output by the last attention module, the edge generator projects each hidden state to a probability distribution over all the possible edge labels:
$\text{P}^{(e)}_i(j) = \:  \text{softmax}(W_e\boldsymbol{h^{d(e)}_i})$
, where $W_e$ are the trainable parameters to map the hidden states to a probability distribution over all the edge labels, and $\text{P}^{(e)}_i(j)$ is the probability of the edge label $j$. 
The one with the highest probability, i.e., $e^d_i = \text{argmax}_j\text{P}^{(e)}_i(j)$,  is predicted as the label of the edge that connects the located parent node and its child node to be generated.

\subsubsection{Node Decoder}
\label{sec:node decoder}
The last step in each decoding iteration is to decode a new child node for the parent node $p^d_i$ with the edge label $e^d_i$. Therefore, the node decoder takes three inputs, including (i) the hidden states output by the previous edge decoder, (ii) the previously-predicted parent node, and (iii) the previously-predicted edge label.
Figure~\ref{fig:architecture} shows that the node decoder contains a parent-edge state fusion layer, followed by three attention modules (which are similar as those in  parent/edge decoders),
and a node generator.

\smallskip\noindent\textbf{Parent-Edge State Fusion.}
\tool{} incorporates a parent-edge state fusion to combine the hidden states of each node $\{\boldsymbol{h^{d(e)}_i}\}$ with the hidden states of its parent node $p^d_i$ and the embedding of the edge $e^d_i$ generated by the edge decoder:
$\boldsymbol{h^{d(n)}_i} = W_f\big(\boldsymbol{h^{d(e)}_i} + \boldsymbol{h^{d(e)}_{p^d_i}} + \boldsymbol{e^d_i}\big)$
, where $\{\boldsymbol{e^d_i}\}$ is the embedding of generated edges $\{e^d_i\}$.

\vspace{0.03in}
\noindent\textbf{Node Generator.}
Similar to the edge generator, the node generator takes the hidden states output by the attention module to compute a probability distribution among all  possible nodes:
$\text{P}^{(n)}_i(j) = \: \text{softmax}(W_n\boldsymbol{h^{d(n)}_i})$
, where $W_n$ is  trainable parameters, and $\text{P}^{(n)}_i(j)$ is the probability of node $j$ being the generated node. The node with the highest probability is generated,
i.e.,  $n^d_i = \text{argmax}_j\text{P}^{(n)}_i(j)$.

\begin{table}[t]
    \footnotesize
    \centering
    \begin{tabular}{c|c|c|c|c}
    \hline
    \multirow[t]{2}{*}{\textbf{Nodes}}&\multicolumn{4}{c}{\textbf{Edge Labels}} \\
    \cline{2-5}
    & \textbf{required} & \textbf{required+} & \textbf{optional} & \textbf{optional*} \\
    \hline \hline
    MemberRef & member & - & qualifier & selectors \\
    FuncInvoc & member & - & qualifier & args \\
    LocalVarDecl & type & declarators & - & - \\
    \hline\hline
    \multicolumn{5}{c}{\textbf{Syntax rules for selecting parents and generating edges}} \\
    \hline\hline
    \multicolumn{5}{l}{}\\
    \multicolumn{5}{l}{
    \multirow{9}{*}{
        \begin{math}
        \begin{aligned}
            \textbf{Rule1:}\: \forall p \in P, \:& \exists e \in E_{req}(p) \cup E_{req+}(p)(F(p,e) = 0) \\
            & \rightarrow p \in P_{must} \land e \in E_{must}(p) \\
            \textbf{Rule2:}\: \forall p \in P, \:& \exists e \in E_{opt}(p) \cup E_{opt*}(p)(F(p,e) = 0) \\
            & \rightarrow p \in P_{might} \land e \in E_{might}(p)\\
            \textbf{Rule3:}\: \forall p \in P, \:& \forall e \in E_{req}(p)(F(p,e) = 1) \:\land 
            E_{req+}(p) = \varnothing \:\land \\
            & \forall e \in E_{opt}(p)(F(p,e) = 1) \:\land 
            E_{opt*}(p) = \varnothing \\
            & \rightarrow p \in P_{invalid} \\
            \textbf{Rule4:}\: \forall p \in P, \:&\forall e \in E(e \notin E_{req} \cup E_{req+} \cup E_{opt} \cup E_{opt*}(p)) \\
            & \rightarrow e \in E_{invalid}(p) \\
        \end{aligned}
        \end{math}
    }} \\
    \multicolumn{5}{l}{}\\
    \multicolumn{5}{l}{}\\
    \multicolumn{5}{l}{}\\
    \multicolumn{5}{l}{}\\
    \multicolumn{5}{l}{}\\
    \multicolumn{5}{l}{}\\
    \multicolumn{5}{l}{}\\
    \multicolumn{5}{l}{}\\
    \multicolumn{5}{l}{}\\
    \multicolumn{5}{l}{}\\
    \multicolumn{5}{l}{}\\
    \hline
    
    \end{tabular}
    \caption{\small{Syntax rules defined in \javalang, based on which \tool designs FOL rules for decoding parents and edges. ``required'' means the node must have one and only one edge with the given label, ``required+'' means the node must have one or more edges with the given label, ``optional'' means the node could have zero or one edge with the given label, and ``optional*'' means the node could have zero, one or multiple edges with the given label.}}
    \label{tab:syntax examples}
\end{table}
\begin{table}[t]
    \centering
    \footnotesize
    \begin{tabular}{c|l}
    \hline
    \textbf{Preconditions} & \multicolumn{1}{c}{\textbf{Semantic \drules for generating nodes}} \\
    \hline
    \hline
    \multirow{5}{*}{\includegraphics[width=0.12\textwidth,height=10.5mm]{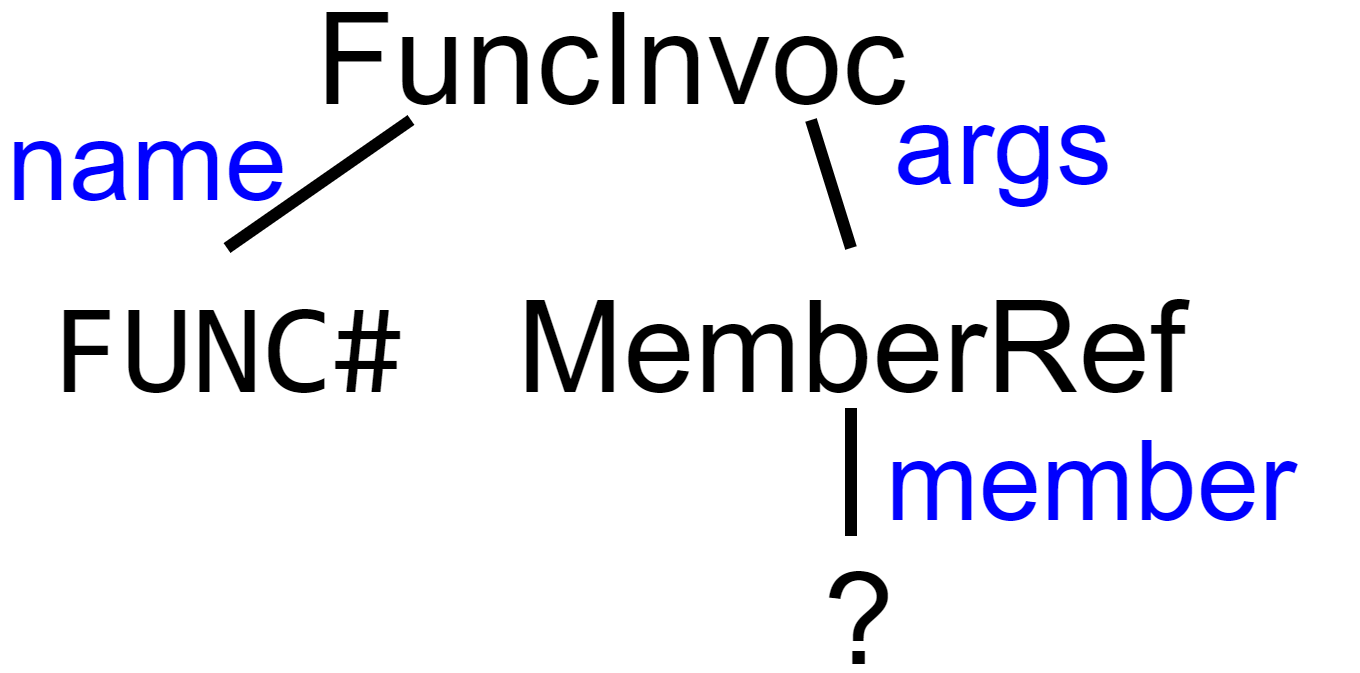}} &
    \multirow{5}{*}{
        \begin{math}
        \begin{aligned}
            \textbf{Rule 5:}\;\forall n \in N, & type(n) \npreceq type(\texttt{FUNC.args}) \\
            & \xrightarrow{} n \in N_{invalid} \\
            \textbf{Rule 6:}\;\forall n \in N, & type(n) \preceq type(\texttt{FUNC.args}) \\
            & \xrightarrow{} n \in N_{might} \\
        \end{aligned}
        \end{math}
    } \\
    & \\
    & \\
    & \\
    & \\
    \hline
    \multirow{5}{*}{\includegraphics[width=0.12\textwidth,height=7.2mm]{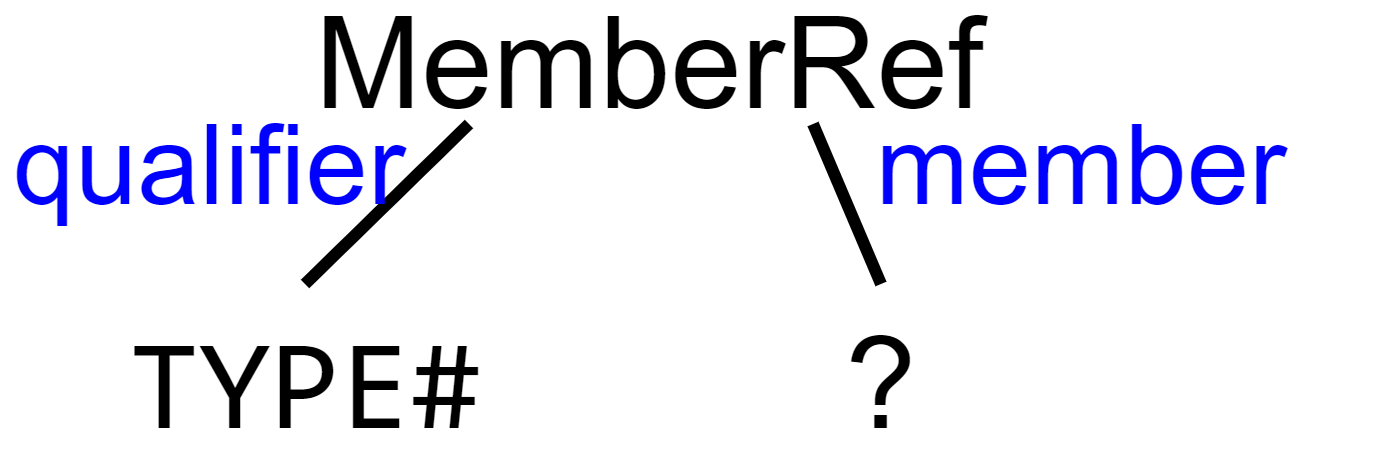}} &
    \multirow{3}{*}{
        \begin{math}
        \begin{aligned}
            \textbf{Rule 7:}\; \forall n \in N, & n \notin \texttt{TYPE.fields} \\
            & \xrightarrow{} n \in N_{invalid} \\
            \textbf{Rule 8:}\; \forall n \in N, & n \in \texttt{TYPE\#.fields} \\
            & \xrightarrow{} n \in N_{might} \\
        \end{aligned}
        \end{math}
    } \\
    & \\
    & \\
    & \\
    & \\
    \hline
    \multirow{5}{*}{\includegraphics[width=0.12\textwidth,height=10.5mm]{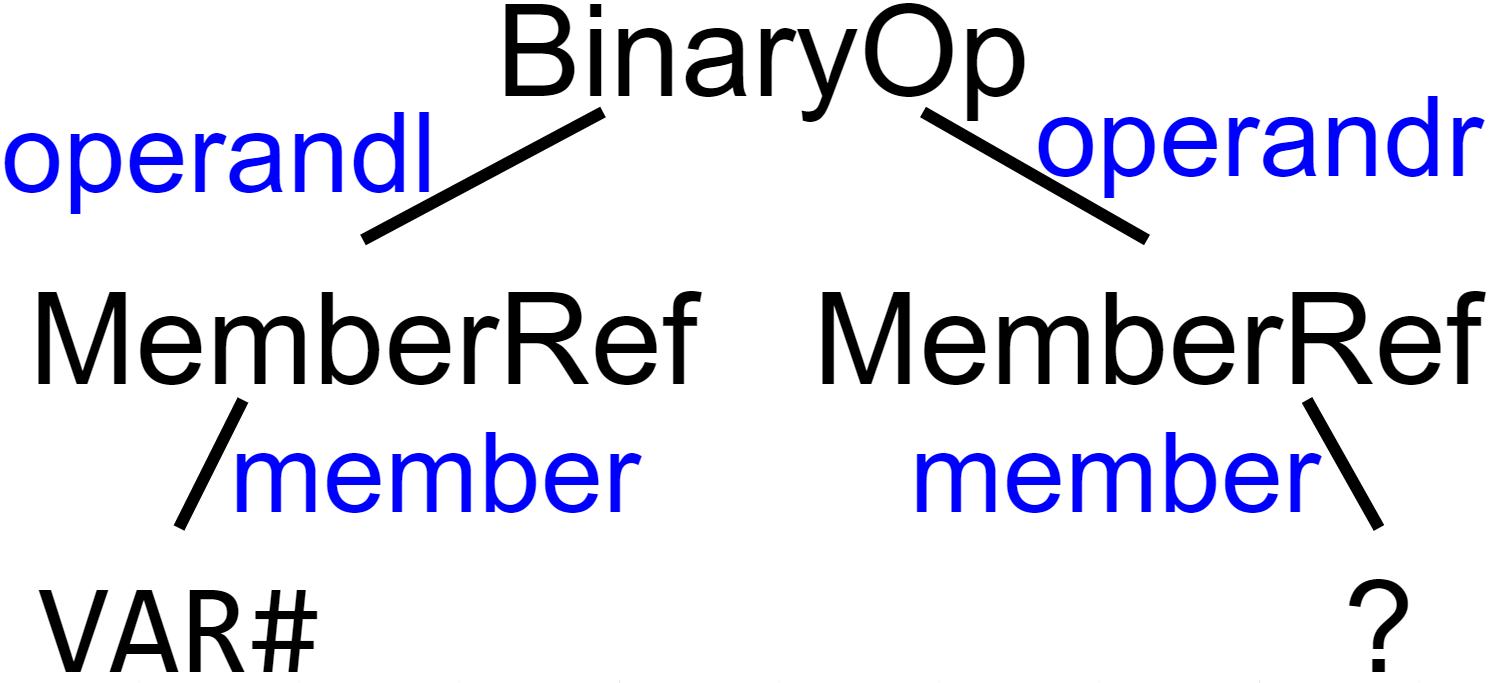}} &
    \multirow{5}{*}{
        \begin{math}
        \begin{aligned}
            \textbf{Rule 9:}\;\forall n \in N, & type(n) \npreceq type(\texttt{VAR\#}) \\
            & \xrightarrow{} n \in N_{invalid} \\
            \textbf{Rule 10:} \forall n \in N&, type(n) \preceq type(\texttt{VAR\#}) \\
            & \xrightarrow{} n \in N_{might}
        \end{aligned}
        \end{math}
    } \\
    & \\
    & \\
    & \\
    & \\
    \hline
    \end{tabular}
    \caption{\small{Examples of \tool's semantic rules designed in \domain module, and preconditions of applying them. Semantic rules are used during generating nodes (\texttt{?} refers to the node to be generated). %The full list is available at~\cite{share}.
    }}
    \label{tab:semantic examples}
%\vspace{-1em}
\end{table}

\begin{figure*}[t]
    \centering
    \includegraphics[width=0.9\linewidth]{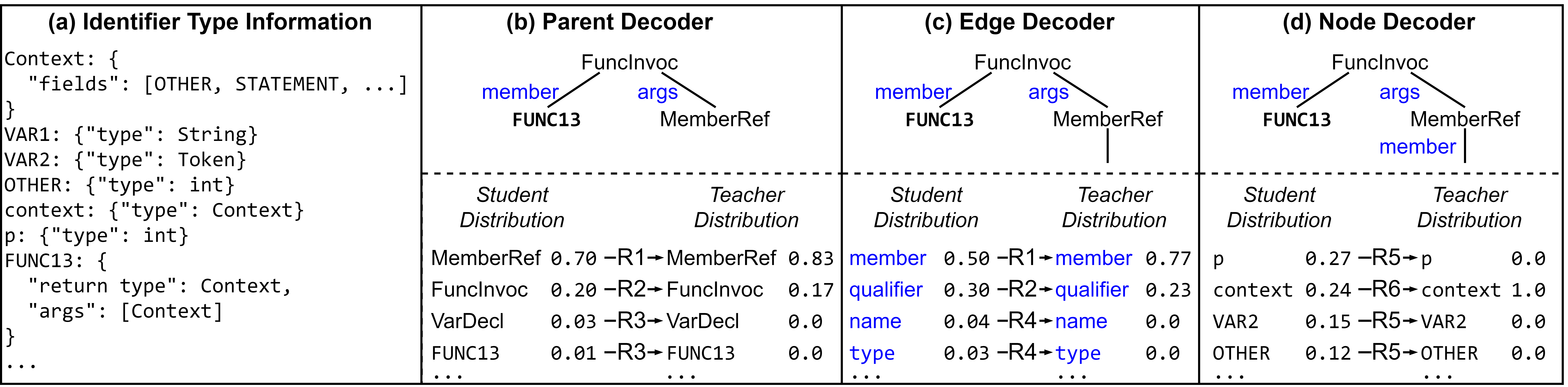}
    \caption{\small Process of \domain modifying the probability distribution and \tool  generating  node \code{context} for \defects{}'s Closure-123. R1 denotes Rule 1. }
    \label{fig:generation example}
\end{figure*}

\subsection{Novelty 2: \Domain}
\label{sec:domain knowledge fusion}

\noindent We propose \domain to force the decoder to generate syntactically and semantically valid patches, 
which (i) represents \emph{syntax and semantics} as {\rules}, (ii) uses \rules to  create \emph{teacher-student distributions}, and (iii) formulates \emph{a new loss function} to guide the models to learn from teacher-student distributions (i.e., to update models' weights) to generate ASTs to follow these \rules during  \emph{both the training and inference phases}. 

\smallskip\noindent\textbf{(i1) Syntactic Rules.} 
Table~\ref{tab:syntax examples} shows  examples of % pre-defined 
domain knowledge for a valid AST in \javalang, where each AST node must  have edges with only a small subset of labels. For example, node \nsty{MemberRef} must have one and  only  one edge of label \esty{member}, meaning that to reference a member, there must be one and only one member (e.g., member \nsty{context} in Figure~\ref{fig:generation example}~\circled{5}).
Such domain knowledge can guide the \decoder to predict correct parents and generate correct edge labels. For example, in a partially generated AST, if node \nsty{MemberRef} has no edge of label \esty{member}, the decoder should predict \nsty{MemberRef}  as parent and generate a \esty{member} edge for it; otherwise, the AST is syntactically wrong. 

To leverage such syntax, we create \rules, shown in Table~\ref{tab:syntax examples}, where $p$ refers to parent node, $P$ is the collection of nodes that might be predicted as a parent by the model, $e$ is edge label, and $E_{req}(p)$, $E_{req+}(p)$, $E_{opt}(p)$, and  $E_{opt*}(p)$ are the sets of edge labels that are required or optional for node $p$ (e.g., $E_{req}(\nsty{MemberRef})=\{\nsty{member}\}$, which is the edge label in the same row as \nsty{MemberRef} and under column \nsty{required}).
$F(p,e)$ is a function that returns the number of edges that link to parent $p$ with label $e$. $P_{must}$, $P_{might}$, and $P_{invalid}$ are the sets of nodes that must be predicted as a parent, might be predicted as a parent, or impossible as a parent in the future decoding iterations. Respectively, $E_{must}(p)$, $E_{might}(p)$, and $E_{invalid}(p)$ are the sets  of edges that must, might, or must not be generated for parent node $p$. 

The intuition of syntactic rules is that any edge labels required by a node must be generated, any edge labels optional for a node could be generated, and if a node has all  potential edges generated, it cannot generate more edges. For example, \textbf{Rule 1} states that for a node with any required edge not generated yet, it must be predicted as parent and the required edge must be generated. 

During every decoding iteration, \domain applies the syntax rules to find parent nodes and corresponding edges belonging to each category (i.e., $P_{must}$, $P_{might}$, etc.), which is  used later (in the (ii) Teacher-Student Distributions step to modify the parent decoder and edge decoder's output probability distribution ($\text{P}^{(p)}_i$ and $\text{P}^{(e)}_i$). 

\noindent\textbf{(i2) Semantic Rules.}
Another important type of domain knowledge---code semantics, i.e., type matching, of which \tool{} further takes advantages, is used to train the model to avoid generating  semantically-invalid ASTs. Specifically, \tool{} uses \javaparser to statically analyze the entire buggy program (not just the buggy function) to collect type information of each accessible identifier. For example, the left of Figure~\ref{fig:generation example} shows the following type information for the Closure-123 bug: (i) the data type for each variable, (ii) the return type and the arguments for each function, and (iii) the fields and the declared functions for each class. With such type information, we design semantic rules for the decoder to generate nodes (Table~\ref{tab:semantic examples}). A rule is applied when its precondition matches the partially generated AST. $n$ refers to leaf node labels, $N$ is the collection of all the accessible identifiers' names, $type(n)$ returns the data type of $n$, and $\preceq$ means type compatibility. $N_{might}$ and $N_{invalid}$ are the collections of nodes that can or cannot be generated. Semantic rules ensure type compatibility during node generation, for example, \textbf{Rule 5} states when the model  generates a member as an argument of a function invocation, only the nodes whose types are compatible (i.e., same or subtype) with the argument type defined in the function's signature are possible to be generated. 

During every decoding iteration, semantic rules are used to find nodes that are possible or invalid to be generated (i.e., nodes belonging to $N_{might}$ and $N_{invalid}$), which is used later (in the (ii) Teacher-Student Distributions step) to modify the node decoder's output probability distribution ($\text{P}^{(n)}_i$). 

\smallskip
\noindent\textbf{(ii) Teacher-Student Distributions.}
To let the APR model learn syntactic and semantic rules, we adapt the teacher-student architecture~\cite{firstlogic}. Specifically, the rules are the teacher's knowledge, which we want the student (the APR model) to learn. The terms teacher and student  follow prior work~\cite{firstlogic}, as the teacher ``teaches'' the model to generate output satisfying the domain knowledge rules that the teacher knows. 

The \emph{student distribution} is the probability distribution output by the \decoder, i.e., $\text{P}^{(p)}$, $\text{P}^{(e)}$ and $\text{P}^{(n)}$. We create the \emph{teacher distribution} $\hat{\text{P}}^{(p)}$ by modifying the student distribution as
\begin{equation}
\small
\hat{\text{P}}^{(p)}_i =
\begin{cases}
1 & p_i \in P_{must}  \\
\text{P}^{(p)}_i & p_i \in P_{might} \\
0 & p_i \in P_{invalid}
\end{cases}
\end{equation}
and re-normalize it, while distributions $\hat{\text{P}}^{(e)}$ and $\hat{\text{P}}^{(n)}$ are created similarly. The teacher distributions satisfy the rules, as the probability of parents, edges, and nodes that must be generated are the highest, and the probability of invalid ones are 0. 

For example, in Figure~\ref{fig:generation example} (b), by applying \textbf{Rule 1}, \nsty{MemberRef} has a required edge \esty{member} not generated and thus belongs to $P_{must}$. The probability of \nsty{MemberRef} is modified to 1, and then re-normalized to 0.83. In Figure~\ref{fig:generation example}(c), after selecting \nsty{MemberRef} as the parent, by applying \textbf{Rules 1, 2 and 4}, \esty{member} is required, \esty{qualifier} is optional, and the rest edges are invalid for \nsty{MemberRef}. Thus, the probability of \esty{member} is set to 1 (re-normalized to 0.77) and the probability of \esty{qualifier} is kept as 0.3 (re-normalized to 0.23). In Figure~\ref{fig:generation example}(d), \domain applies \textbf{Rules 5 and 6} as the partially generated AST matches their preconditions.
Thus, we keep only identifiers (e.g., \code{context}) that are types compatible with the argument type of \code{FUNC13} (i.e., type \code{Context} is valid). 
Our \domain sets the  probabilities of all other identifiers with incompatible types to 0. As a result, the decoder generates the correct node \code{context} instead of \code{p} of type \code{int} that has the highest probability in the student distribution.

\smallskip
\noindent\textbf{(iii) Distillation.}
Distillation transfers the domain knowledge formulated by syntax and semantic rules into the APR models' weights~\cite{firstlogic}, which speeds up the training and helps  APR models learn the domain knowledge and compute better probability distributions. Distillation is performed by introducing an extra loss to minimize the difference between student and teacher distribution. The overall training objective is to minimize  
(i) the loss between the student distributions and the ground-truth labels, and (ii) the loss between the student distributions and the teacher distributions, where (iii) is the standard loss used by APR techniques~\cite{jiang2021cure,lutellier2020coconut,zhu2021recoder,chen2018sequencer}, while (2) is a novel contribution of this paper. Specifically, the joint loss is as follows: 

\begin{equation}
\footnotesize
\begin{aligned}
    Loss = & L_{CE}(\text{P}^{(p)}, y^{(p)}) + L_{CE}(\text{P}^{(e)}, y^{(e)}) + L_{CE}(\text{P}^{(n)}, y^{(n)}) \\
    & L_{KL}(\text{P}^{(p)}, \hat{\text{P}}^{(p)}) + L_{KL}(\text{P}^{(e)}, \hat{\text{P}}^{(e)}) + L_{KL}(\text{P}^{(n)}, \hat{\text{P}}^{(n)})
\end{aligned}
\end{equation}
where the $L_{CE}()s$ are cross-entropy~\cite{crossentropy} between the student distribution and the ground-truth labels ($y^{(p)}$, $y^{(e)}$ and $y^{(n)}$ are the ground-truth for parents, edges, and nodes respectively), and $L_{KL}()$s are the  loss that we add, which  calculates the Kullback–Leibler divergence~\cite{kldivergence} between the student distributions and the teacher distributions.
By minimizing the training objective during training, the APR models learn to generate syntactically and semantically correct ASTs. 

\subsection{Training and Inference}
\label{sec:trainandinter}

\noindent In the training phase, the model takes the \asg{} of the normalized buggy function and AST of normalized patched code to learn the transformation from the former to the latter. We also leverage ensemble learning~\cite{ensemble1,ensemble2} to train multiple models to increase the diversity of the learned fix patterns. Following previous work~\cite{lutellier2020coconut,jiang2021cure}, we first train different models with random hyper-parameters (e.g., number of encoder blocks, decoder blocks, or hidden states dimension), and then select the Top-$k$ models according to their loss on the validation set.

In the inference phase, for the given buggy function, each of the $k$ trained models generates a list of ASTs of normalized patched code. Among all the generated ASTs, \tool{} first ranks them via their ranks and average probabilities.

\subsection{Patch Generation and Validation} 
\label{sec:patch generation}

\noindent The generated ASTs are converted to normalized source code, which still contains normalized tokens. \tool{} reconstructs them into concrete patches by replacing the normalized tokens with the corresponding concrete identifiers, based on the mapping recorded in the code normalization phase (Section~\ref{sec:data representation}).
For example, in Figures~\ref{fig:ast example}~\circled{6} and \circled{7}, the normalized tokens \code{VAR3} and \code{FUNC13} are replaced by the concrete identifiers \code{rhsContext} and \code{getContextForNoInOperator}, respectively. Such concrete patched code are the final patches generated by \tool{}. For normalized patched code with \emph{unknown} tokens (e.g., \code{TYPE-UNK}), \tool performs type-analysis to find compatible concrete values for reconstruction. In particular,
for each unknown token, \tool first analyzes its parent and sibling (if any) nodes in the ASTs, summarize rules that the unknown token should follow (e.g., its data type), and then replaces the unknown tokens with all valid concrete identifiers to ensure semantic matching. 
%The reconstruction of  unknown tokens ensures semantic matching. 
%For example, Figure~\ref{fig:unique patch}(b) presents the correct patch generated by \tool{} for the bug Closure-5. In particular, \tool{} first generates an AST, which is converted to normalized code ``\code{if (gramps.FUNC-UNK())\{return false;\}}''. \tool then performs type-analysis to find that \code{FUNC-UNK} should be a function that is declared in variable \code{gramps} and returns a boolean value. Thus, \tool{} replaces \code{FUNC-UNK} with all the accessible function names that satisfy the constraints, and the correct patch is included among these concretized patches. %\nan{updated}

All  generated patches are  validated against test cases. Following prior work~\cite{lutellier2020coconut,yang2017validate,jiang2021cure},  validation terminates when it finds a \emph{plausible} patch that either (i) passes all  test cases or (ii) passes all  originally-passed tests and at least one originally-failed test case.

\section{Experiment Setup}
\label{experiment setup}

\noindent We evaluate \tool{}  with three research questions: \textbf{RQ1: Effectiveness and Generalizability.} How does \tool{} perform compared to existing APR techniques? \textbf{RQ2: Ablation Study.} What is the contribution of each component in \tool? and \textbf{RQ3: Ranking.} How does \tool{} rank the correct patches compared to other tools? 
% How does \tool{} perform on the other bug benchmarks?

\subsection{Datasets}
\noindent\textbf{Training Data.} 
We construct the training data for \tool{} from the dataset shared in previous work~\cite{lutellier2020coconut,zhu2021recoder}, which are mined from open-source GitHub Java projects.
%We construct the training data for \tool{} from two sources: dataset mined from open-source GitHub Java projects and the existing dataset released in the recent DL-based APR work~\cite{zhu2021recoder}. The existing dataset contains 103,585 pairs of  buggy programs and their developer patches. For the new dataset, we collect the commits before 2006 containing  keywords ``fixed'', ``bug'', and ``patch'' from GitHub Java projects. 
Following previous work~\cite{zhu2021recoder,jiang2021cure,lutellier2020coconut}, we remove projects that are in or cloned from \defects{} projects from our training set.
In total, our training data contains 576,002 pairs of buggy programs and their developer patches, which is randomly split into training set (90\%) and validation set (10\%). The validation set is used to tune and select models.

\noindent\textbf{Bug Benchmarks.} 
We evaluate \tool{} on  three well-established bug benchmarks, including: (1) \defectsold{}~\cite{defects4j}, the most widely-used version of the Defects4J benchmark with 393 Java  bugs~\footnote{Following previous work~\cite{jiang2021cure,lutellier2020coconut}, two duplicated bugs (Closure-63 and 93) are removed from our evaluation.}, (2) \defectsnew{}~\cite{defects4j}: the latest version of the Defects4J benchmark with additional 444 Java bugs, and (3) \quixbugs{}~\cite{quixbugs}, the widely-used benchmark with 40 Java bugs.

\subsection{Evaluated Techniques}

\noindent To compare the effectiveness of \tool{} with existing APR techniques, we include the following state-of-the-art APR techniques for comparison.

%Following previous work~\cite{jiang2021cure,lutellier2020coconut,perfectloc,avatar}, we compare the number of bugs correctly fixed by each technique with the perfect fault localization to mitigate the impact of different fault localization techniques~\cite{perfectloc}.

\begin{itemize}
    % \item \textbf{Non-DL-based APR:} we compare \tool{} with SimFix~\cite{simfix}, TBar~\cite{tbar}, and ACS~\cite{ACS}, since they are the most effective non-DL-based APR techniques~\cite{efficiency} in the heuristic-based, template-based, and constraint-based categories, respectively. 
    \item \textbf{Non-DL-based APR:} we compare \tool{} with SimFix~\cite{simfix} and TBar~\cite{tbar}, since they are the most effective heuristic-based and template-based non-DL-based APR techniques~\cite{efficiency}.
    
    \item \textbf{DL-based APR:} we compare \tool{} with state-of-the-art DL-based APR techniques for Java programs, including \sequencer{}~\cite{chen2018sequencer}, 
    %\codit{}~\cite{codit}, 
    \dlfix{}~\cite{Li2020dlfix}, \coconut{}~\cite{lutellier2020coconut}, \cure{}~\cite{jiang2021cure},  \rewardrepair{} \cite{rewardrepair}, and \recoder{}~\cite{zhu2021recoder}.
\end{itemize}

\noindent To study the contribution of each novelty of \tool{}, we implement and evaluate the following variants of \tool{}.

\begin{itemize}
    \item \textbf{\tooldecoder{}}: replacing the entire \decoder{} with a traditional sequential decoder. 
    \item \textbf{\tooltrain{}}: removing  \domain{} from the decoder during training (keeping it during  inference).
    \item \textbf{\toolinf{}}: removing  \domain{} from the decoder during inference (keeping it during training).
\end{itemize}

\subsection{Experimental Procedure}
\noindent\textbf{Fault Localization.}
We perform our experiment under two different settings of fault localization: (1) perfect localization, where the actually fault localization is given to the tools, and (2) spectrum based fault localization, where \tool{} uses the suspicious faulty locations reported by Ochiai~\cite{ochiai} (a spectrum based fault localization tool). Both settings are widely used in previous works~\cite{rewardrepair,zhu2021recoder,jiang2021cure,lutellier2020coconut}.

\smallskip\noindent\textbf{Patch Correctness.} 
In line with previous work~\cite{jiang2021cure,lutellier2020coconut,rewardrepair,zhu2021recoder}, for evaluation purpose only, we manually check the correctness of plausible patches returned by \tool{}. We consider a plausible patch \emph{correct} if it is semantically equivalent to developer patches. The  labeling procedure involves two participants. The agreement ratio is 92.1\% and inconsistent cases are resolved by further discussion. 
%Our labelled results are publicly available at ~\cite{}.  

\smallskip\noindent\textbf{Implementation.} 
For \asg{} construction, we use the widely-used toolkits \javalang~\cite{2019javalang} and \javaparser~\cite{smith2019javaparser} to first parse buggy functions and patches into ASTs. The APR models are implemented with PyTorch~\cite{pytorch}. To select the hyperparameters, we use random search within the following range: number of encoder layers (6-8), number of parent and edge decoder layers (1-2), number of node decoder layers (4-8), embedding and hidden states dimension (256-384). 
%, number of attention heads (4-8)
We use a dropout rate set to 0.1 to avoid overfitting, and use Adam optimizer with learning rate being $2.5e^{-4}$.
%and is adjusted by a cosine scheduler~\cite{cosine} with warm-up of 2,000 training steps. 
%Following previous work~\cite{lutellier2020coconut,jiang2021cure}, we apply ensemble technique and 
We tune the top-5 models with the lowest perplexity on the validation set until convergence for ensemble learning.
In the inference stage, we use beam search~\cite{beamsearch} with beam size set to 1,000 to generate patches for each bug in the bug benchmarks.  During validation, we set a five-hour running-time limit, which is the same as existing work~\cite{simfix,Li2020dlfix,lutellier2020coconut,hercules,zhu2021recoder}.

\smallskip\noindent\textbf{Infrastructure.}
We train \tool on one 56-core server with eight NVIDIA GeForce RTX 2080 TI GPUs, \rev{and evaluate \tool on the same server with one NVIDIA GeForce RTX 2080 TI GPU.}

\subsection{Threats to Validity}
\noindent \emph{Threats to internal validity} lie in the approach implementation and manual patch correctness identification. To mitigate these threats, multiple authors check the code and participate in the manual labeling procedure. \emph{Threats to external validity} lie in bug benchmarks used in our evaluation, which cannot guarantee the generalizability on other benchmarks. To mitigate these threats, we perform our experiments on three widely-used benchmarks with up to 877 real-world Java bugs. Evaluation on more benchmarks of different program languages could be done in the future since our approach is not specifically designed for Java.

\section{Result}
\label{sec:result}

\subsection{RQ1: Effectiveness and Generalizability}
\label{sec:rq1}

\begin{table}[htb]
	\centering
	\small
	\begin{adjustbox}{width=0.95\columnwidth}
	\begin{tabular}{lccc}
		\hline
        \textbf{Techniques} & \textbf{\defectsold} &  \textbf{\defectsnew} & \textbf{\quixbugs} \\
        \hline\hline
        \sequencer{}~\cite{chen2018sequencer} & 14 & - & -\\
        % ACS~\cite{ACS} & 16 & - & - \\ 
        % \codit{}~\cite{codit} & 16 & - & - \\
        SimFix~\cite{simfix} & 28 & - & -\\
        \dlfix{}~\cite{Li2020dlfix} & 38 & - & -\\
        \coconut{}~\cite{lutellier2020coconut} & 44 & - & 13 \\
        \rewardrepair{}~\cite{rewardrepair} & 45 & 45 & 20 \\
        TBar~\cite{tbar} & 53 & - & - \\
        \cure{}~\cite{jiang2021cure} & 56 & 19 & \textbf{25} \\
        \recoder{}~\cite{zhu2021recoder} & 64 & - & 17 \\ 
        \hline\hline
        \textbf{\tool{} (our approach)} & \textbf{72} & \textbf{50} & \textbf{25} \\
	    \hline
	\end{tabular}
	\end{adjustbox}
	\smallskip
	\caption{\small{Number of correctly fixed bugs by each tool on three benchmarks with perfect fault localization. 
	%Number with ``$*$'' is obtained with spectrum based fault localization~\cite{ochiai}, while the others are with perfect fault localization. 
	``-'' means that tool has not released its performance on the benchmark.
	}}
	\label{table:effectiveness-pfl}
\end{table}

\begin{table}[htb]
	\centering
	\small
	\begin{adjustbox}{width=0.95\columnwidth}
	\begin{tabular}{lccc}
		\hline
        \textbf{Techniques} & \textbf{\defectsold} &  \textbf{\defectsnew} & \textbf{\quixbugs} \\
        \hline\hline
        % \sequencer{}~\cite{chen2018sequencer} & 14 & - & -\\
        % \coconut{}~\cite{lutellier2020coconut} & 44 & - & 13 \\
        TBar~\cite{tbar} & 23 & 8 & - \\
        SimFix~\cite{simfix} & 24 & 2 & -\\
        % \cure{}~\cite{jiang2021cure} & - & - &  \\
        \rewardrepair{}~\cite{rewardrepair} & 29 & 22 & 19 \\
        \dlfix{}~\cite{Li2020dlfix} & 30 & - & -\\
        \recoder{}~\cite{zhu2021recoder} & \textbf{45} & 19 & 17 \\ 
        \hline\hline
        \textbf{\tool{} (our approach)} & 38 & \textbf{24} & \textbf{23} \\
	    \hline
	\end{tabular}
	\end{adjustbox}
	\smallskip
	\caption{\small{Number of correctly fixed bugs by each tool on three benchmarks with spectrum based fault localization~\cite{ochiai}.
	}}
	\label{table:effectiveness-nfl}
\end{table}

\noindent \textbf{Results with perfect fault localization.}
Table~\ref{table:effectiveness-pfl} shows the number of bugs that are correctly fixed by \tool{} and other APR techniques on three benchmarks with perfect fault localization. \tool{} fixes 72 bugs, outperforming all the compared techniques on the most widely-used benchmark \defectsold{}, 8 and 19 more bugs than the best DL-based and non-DL-based APR approach \recoder{} and \tbar{} respectively. In addition to the widely-used benchmark \defectsold{}, Table~\ref{table:effectiveness-pfl} also presents the effectiveness of \tool{} on additional two benchmarks, \defectsnew{} and \quixbugs{}. 
\tool{} is consistently effective on both additional benchmarks, i.e., fixing 50 bugs on \defectsnew{} and 25 bugs on \quixbugs{}, indicating the generalizability of \tool{} on different bugs. 

In addition, we calculate the patch precision of \tool{}, i.e., the ratio of correct patches to plausible patches. We find that \tool{} achieves 86.7\% precision (i.e., 72 out of 83 plausible patches generated for \defectsold{} are correct), which is substantially higher than existing APR techniques under the same configuration~\cite{jiang2021cure} (e.g., the top-3 precision of existing APR \dlfix{}/\tbar{}/\rewardrepair{} is 58.4\%/62.4\%/70.3\%).

\begin{figure}
    \centering
    \includegraphics[width=0.9\linewidth]{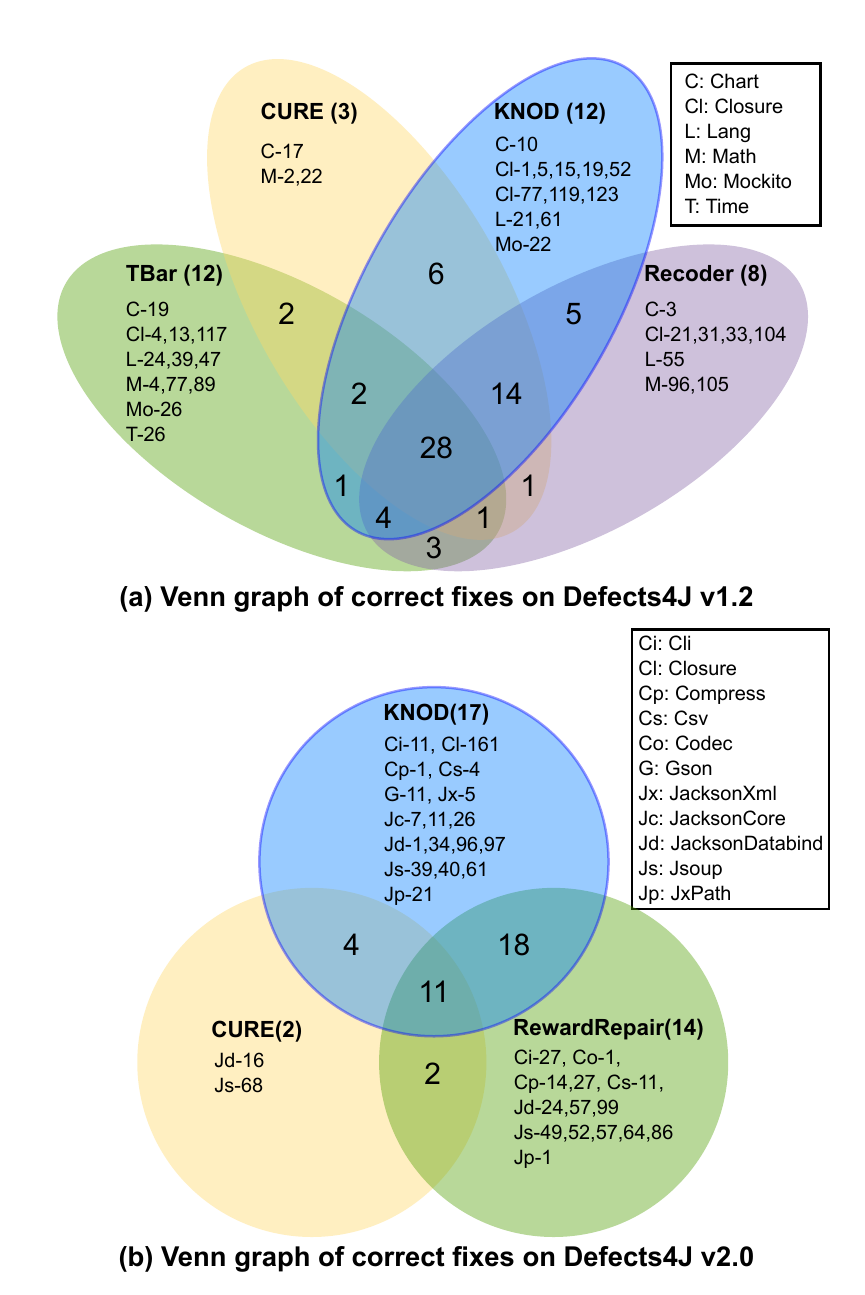}
    \caption{Uniquely fixed bugs  of \tool and the existing state-of-the-art tools with perfect fault localization. Numbers in () are the number of uniquely fixed bugs. 
    %\lin{swap KNOD and CURE color? Blue stands out more. Make TBar and Recoder color a bit more dull}
    }
    \label{fig:venn}
\end{figure}

\smallskip
\noindent \textbf{Results with spectrum-based fault localization.} Table~\ref{table:effectiveness-nfl} shows the number of bugs correctly fixed by~\tool{} and other tools with spectrum-based fault localization. \tool{} still fixes the most number of bugs on~\defectsnew{} and~\quixbugs{}, 24 and 23 respectively. \tool{} fixes the second most on~\defectsold{} (38) and still outperforms the most recent APR paper RewardRepair (29). %, only behind~\recoder{} (45). 
By analyzing the bugs that~\recoder{}  fixes but~\tool{} does not, we find~\tool{}  correctly fixes most of them with perfect localization,  suggesting that \tool{} can achieve better results with a better fault localization technique.

\begin{figure}[h]
    \centering
    \includegraphics[width=0.49\textwidth]{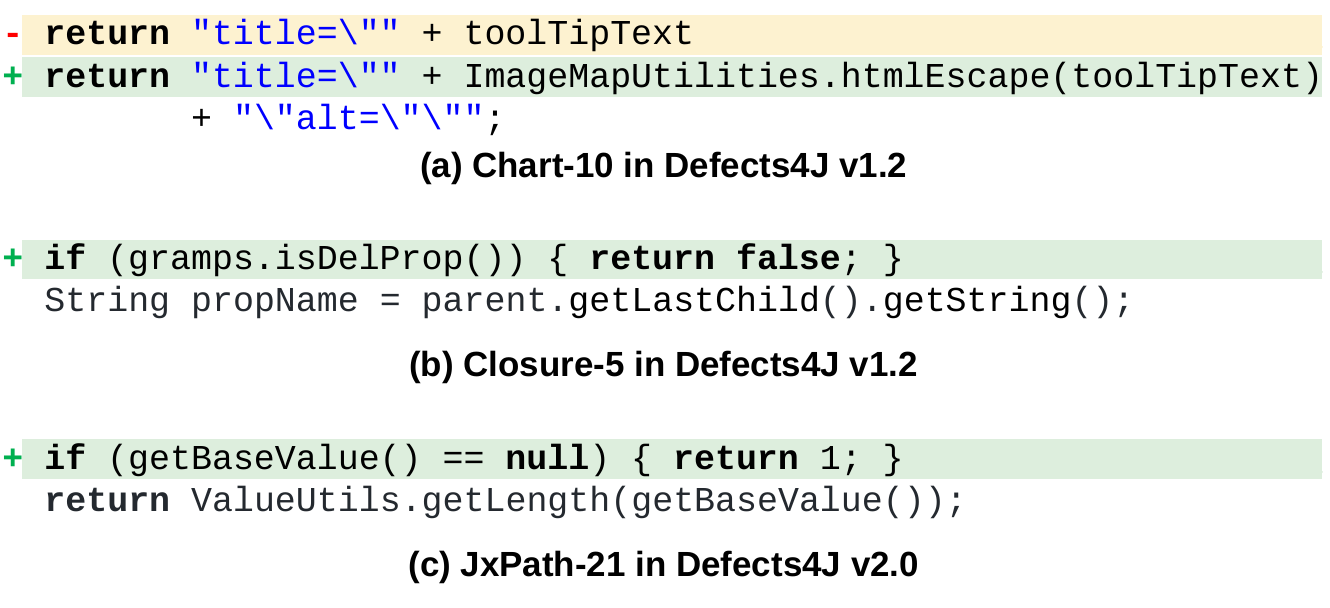}
    \caption{Examples of bugs only fixed by \tool.}
    \label{fig:unique patch}
\end{figure}

\smallskip
\noindent\textbf{Uniquely Fixed Bugs.} 
Figure~\ref{fig:venn}(a) presents the number of overlapped and unique bugs that are fixed by \tool{} and the other three APR techniques that fix the most number of bugs on \defectsold{} (i.e., \tbar{}, \cure{}, and \recoder{}). \tool{} complements the state-of-the-art APR techniques by fixing 12 unique bugs.
On~\defectsnew{}, \tool{} complements~\rewardrepair{} by fixing 21 unique bugs, and complements~\cure{} by fixing 35 unique bugs (as shown in Figure~\ref{fig:venn}(b)), which shows \tool{} can be an excellent complementary technique to existing APR tools. 

Figure~\ref{fig:unique patch} presents three bugs---Chart-10, Closure-5, and JxPath-21---that are fixed only by \tool{}. \tool{} is able to fix Chart-10, which is hard to fix as it involves two project-specific identifiers (\code{ImageMapUtilities} and \code{htmlEscape}) that are declared outside of the buggy function, \rev{thanks to the \domain that helps find the correct type-matched  function call.} \tool{} is also good at inserting code snippets to fix bugs (Figure~\ref{fig:unique patch} (b) and (c)).

\smallskip
\noindent\textbf{Execution Time.} \tool{} spends 12.8s on average generating one thousand candidate patches for a given bug (using one NVIDIA RTX 2080 TI GPU). 

\subsection{RQ2: Ablation Study}
\label{sec:rq2}

\noindent Table~\ref{table:ablation} shows the effectiveness of each novel component of \tool  on \defectsold{}: adding each component improves the effectiveness of \tool{}.
Specifically, the novel \decoder{} enables \tool to fix 16 more bugs (\tool{} vs. \tooldecoder), as sequential decoder cannot leverage structural information well. Adding \domain during training fixes 10 more bugs (\tool{} vs. \tooltrain{}), which means only applying syntax and semantic checking in inference do not work well as the model gives poor ranking without learning \domain during training.
Yet, including \domain{} in the inference phase also help to fix 3 more bugs (\tool vs. \toolinf{}), which can be considered as a second guarantee of syntax/semantic checking. 

\rev{In addition to the number of correct fixes, Table~\ref{table:ablation} also includes the compilation rate of patches generated by each model. Without the \decoder{}, \tooldecoder{} generates a lot more uncompilable patches, which shows that the sequential decoder fails to learn code syntax and semantics well. Moreover, applying \domain{} during training and inference stages both help with generating more compilable patches.}

In summary, each component of \tool{} positively contributes to its effectiveness. The superiority of \toolinf{} to  \tooltrain{} also confirms that injecting \domain into the \emph{training} phase is more effective than only considering domain knowledge in the \emph{inference} phase. 

\begin{table}[htb]
    \centering
    \begin{tabular}{l|cc}
    \hline
        \textbf{Variants} & \textbf{\#Bugs} & \textbf{Compilation Rate}  \\
    \hline \hline
        \tooldecoder & 56 & 33.6\% \\ 
        \tooltrain & 62 & 43.8\% \\
        \toolinf & 69 & 46.1\% \\
        \textbf{\tool{}} & \textbf{72} & \textbf{47.0\%} \\
    \hline
    \end{tabular}
    \caption{Ablation study of each component on \defectsold{}.}
    \label{table:ablation}
\end{table}

\begin{figure}[h]
    \centering
    \subfigure[Ranking of correct fixes on \defectsold{}]{
        \includegraphics[width=0.48\textwidth]{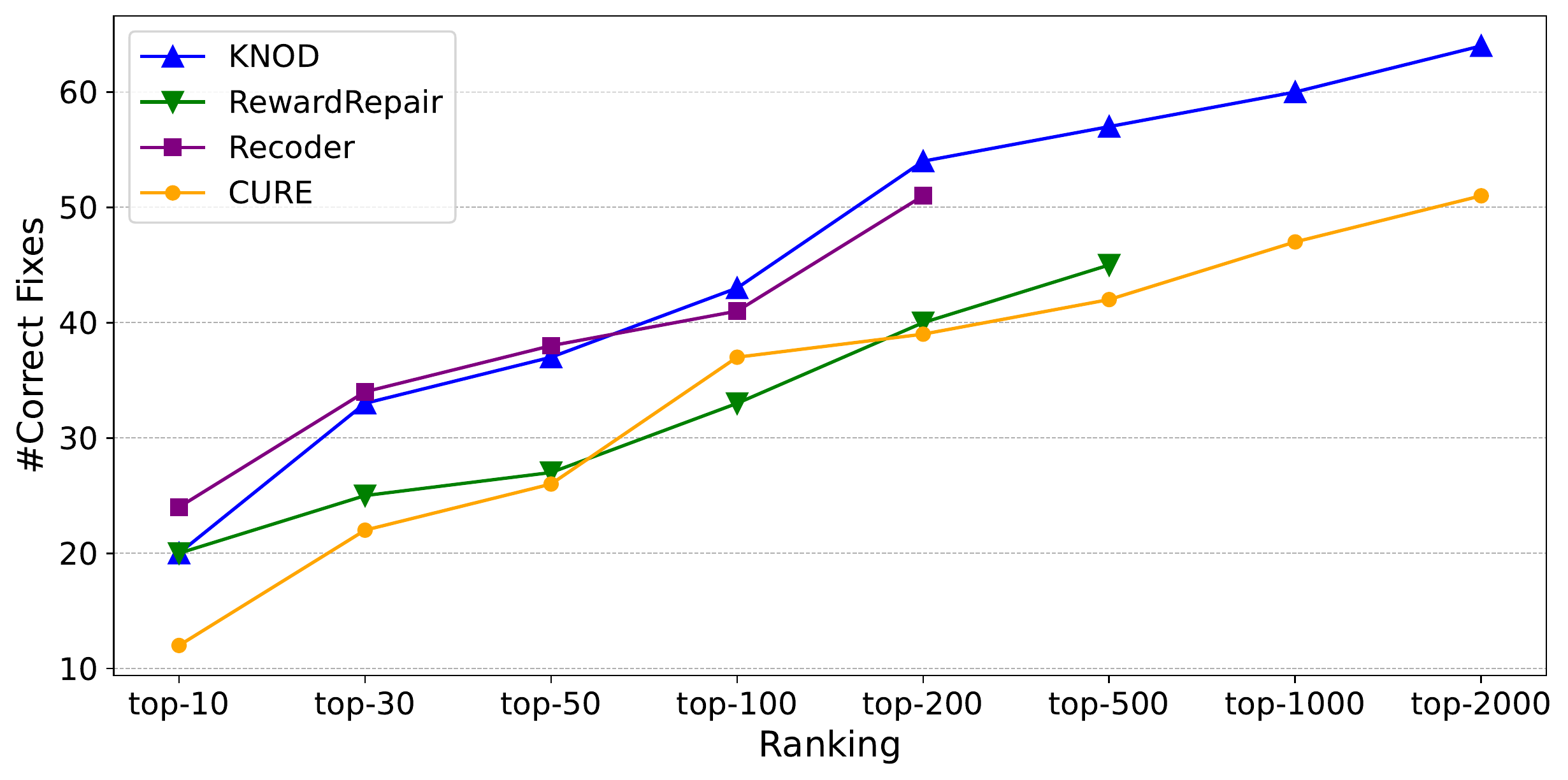}
    }
    \subfigure[Ranking of correct fixes on \defectsnew]{
        \includegraphics[width=0.48\textwidth]{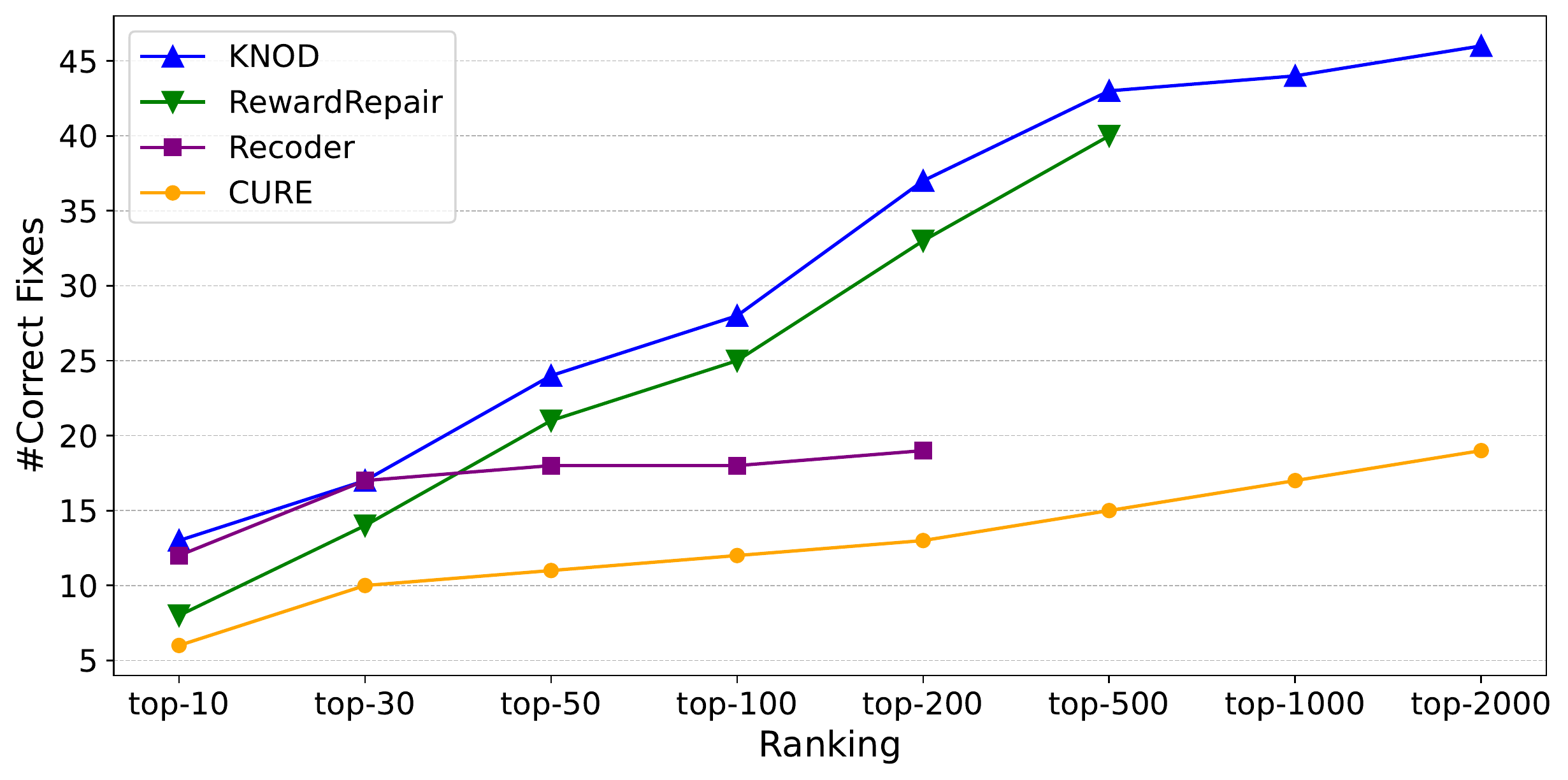}
    }
    \caption{Ranking of correct fixes generated by \tool{} and existing state-of-the-art DL-based APR tools. %\lin{pls reverse the order of the legent: KNOD on the top, ... }\nan{updated}
    }
    \label{fig:rank}
\end{figure}

\subsection{RQ3: Ranking}
\rev{\noindent We further study the ranking of the correct fixes generated by \tool{}.
In Figure~\ref{fig:rank}, we compare \tool{}'s ranking with those of \cure{}, \recoder{} and \rewardrepair{}  on \defectsold{} and \defectsnew{}. Other tools were excluded because they either have not shared the list of the generated candidate patches, or we cannot successfully reproduce their results. For a fairer comparison, we set the beam size to 200 in this experiment, which is the smallest beam size used by \cure{}, \recoder{} and \rewardrepair{}.
KNOD uses the ensemble result from five models, while CURE uses a ensemble of ten models and the other two techniques use one. Since the number of models used is a design choice of different techniques, we compare these techniques as is for a fair ranking comparison.

\tool{} generates more correct fixes than \cure and \rewardrepair{} in all the ranges from top-10 to top-2000, which shows that \tool{} consistently outperforms \rewardrepair{} and \cure{}. Although \recoder{} fixes a competitive number of bugs as \tool{} on \defectsold{}, \tool{} generates a lot more correct fixes on \defectsnew{} in the top-200 patches (37 versus 19), which shows \tool{}'s better generalizability to other benchmarks.}

Since one can choose the number of candidate patches to validate based on the resource budget, our results show that overall, \tool{} fixes the most number of bugs in all budget settings. 

\section{Limitation}

\noindent The main limitations of \tool{} are that (1) \tool{} cannot fix multi-hunk bugs (i.e., bugs that need fix for multiple locations and files) very well (although \tool{} can fix some relatively simple multi-hunk bugs), and (2) the performance of \tool{} depends on the accuracy of fault localization tools. These limitations are shared by most DL-based APR techniques. An important bottleneck in fixing multi-hunk bugs is fault localization, which is an orthogonal problem, which means fixing multi-hunk bugs and developing better fault localization tools are important related works to explore. Despite the limitations, \tool{} still outperforms the existing state-of-the-art APR tools on many benchmarks.
\section{Related Work}
\label{sec:related}

\subsection{DL-based APR}
\noindent Researchers have proposed various APR  techniques~\cite{DBLP:journals/cacm/GouesPR19,DBLP:journals/csur/Monperrus18,livingapr}, which leverage heuristics~\cite{DBLP:journals/tse/YuanB20/arja,DBLP:conf/icse/WenCWHC18,DBLP:conf/kbse/SahaLYP17/elixir}, templates~\cite{tbar,DBLP:conf/issta/GhanbariBZ19}, constraints~\cite{DBLP:conf/icse/XiongWYZH0017/acs, DBLP:journals/tse/XuanMDCMDBM17/nopol,DBLP:conf/ssbse/MartinezM18/cardumen}, or advanced deep learning techniques~\cite{rewardrepair, chen2018sequencer, lutellier2020coconut,jiang2021cure,hopity, zhu2021recoder,deepdebug} to enable patch generation. 
% Our work belongs to DL-based APR, which treats the program repair as the neural machine translation from the buggy code to the fixed code.
Based on the output formats of the decoders, existing DL-based APR can be categorized into \emph{code-generation APR}~\cite{chen2018sequencer, lutellier2020coconut, jiang2021cure} or \emph{edit-generation APR}~\cite{DBLP:conf/icse/TarlowMRCMSA20,zhu2021recoder,hopity}. In particular, the decoders of edit-generation APR~\cite{zhu2021recoder, hopity} first generate edits at different levels (e.g., AST-level edits) and then transform the buggy code into patches based on these edits; the code-generation APR tools,  \sequencer~\cite{chen2018sequencer}, \coconut{}~\cite{lutellier2020coconut}, \cure{}~\cite{jiang2021cure} and \rewardrepair{}~\cite{rewardrepair}, directly generate the code sequences of patches; while \dlfix{}~\cite{Li2020dlfix}, \codit{}~\cite{codit}, and our approach \tool{} generate the ASTs of patches. Different from existing work that generates AST or AST edits based on production rules, \tool{} directly generates ASTs with an explicit tree structure, which forces the model to capture the code structure. To this end, our work proposes a novel three-stage decoder with domain-rule distillation to comprehensively utilize domain knowledge in source code. 

% The only existing APR that incorporates domain knowledge into the training phase (\rewardrepair~\cite{rewardrepair}) 
\rewardrepair{}~\cite{rewardrepair} trains models based on \emph{dynamic} domain knowledge (i.e., the patch execution and compilation information). \tool is different because it leverages \emph{static} domain knowledge, since collecting such dynamic domain knowledge and incorporating it into the training phases is often very expensive. 
Also, \tool uses direct and finer-grained syntactic and type rules (as opposed to indirect and coarser-grained test case passing/failing information). 
% In addition, \tool uses teacher-student distributions for distillation and  designs a novel three-stage tree decoder as opposed to \rewardrepair's sequence decoder. 

\subsection{DL-based Code Generation}
\noindent Recent code generation techniques leverage advanced DL to directly generate code from natural language specifications. Early DL-based code generation techniques~\cite{DBLP:conf/acl/LingBGHKWS16} generate code based on tokens. Due to the rich structural information in source code, recent work~\cite{DBLP:conf/acl/DongL16, DBLP:conf/acl/YinN17, DBLP:conf/emnlp/YinN18, DBLP:conf/acl/RabinovichSK17, DBLP:conf/aaai/SunZXSMZ20} leverages encoder-decoder architectures to generate ASTs. Different from DL-based code generation, our technique is designed for program repair and our proposed decoder is novel in architecture and domain-rule distillation.  In addition to code generation, some DL-based techniques~\cite{DBLP:conf/iclr/YaoXYSN21, DBLP:conf/iclr/YinNABG19,DBLP:journals/pacmpl/Brody0Y20} generate token-level or AST-level edits for program. Instead of generating edits, our approach directly generates patch code in the AST format via a novel decoder. 

\rev{A recent direction of DL-based code generation is applying large language models (LLMs) trained on source code to generate code, such as CodeBert~\cite{codebert}, CodeT5~\cite{codet5}, CodeGen~\cite{codegen}, InCoder~\cite{incoder}, and % OpenAI's 
Codex~\cite{codex}. These LLMs are generic models, while \tool is a customized model that contains a novel three-stage tree decoder and domain-knowledge distillation to fix bugs.}
%Despite their strong ability in code generation, these LLMs are not designed for repairing, and  cannot leverage domain knowledge such as syntactical and semantic rules in code. Making LLMs learn domain knowledge remains future work.}

\section{Conclusion}
\noindent We propose a DL-based APR approach \tool{}, which incorporates domain knowledge to guide patch generation in a \emph{direct and comprehensive} way. \tool{} includes (1) a novel {three-stage decoder} to \emph{directly} generate patch ASTs based on the inherent tree structure of ASTs with \emph{three decoders}, and (2) a novel {domain-rule distillation} component to explicitly inject domain knowledge into the decoding procedure during \emph{both the training and inference phases}. \tool{} is consistently effective on three widely-used benchmarks, fixing 147 bugs in total with perfect fault localization. Our ablation study further confirms the contribution of both novelties in our domain-knowledge-guided tree-decoder architecture.

\smallskip \noindent \textbf{Data Availability:} our replication package is available at~\cite{share}.

\section*{Acknowledgment}
\rev{\noindent We thank the reviewers for their insightful comments and suggestions. This work is partially supported by a J.P. Morgan AI Faculty Research Award. Any opinions, findings, and conclusions in this paper are those of the authors only and do not necessarily reflect the views of our sponsors.}

\bibliographystyle{./bibliography/IEEEtran}
\bibliography{paper}

% Generated by IEEEtran.bst, version: 1.12 (2007/01/11)
\begin{thebibliography}{10}
\providecommand{\url}[1]{#1}
\csname url@samestyle\endcsname
\providecommand{\newblock}{\relax}
\providecommand{\bibinfo}[2]{#2}
\providecommand{\BIBentrySTDinterwordspacing}{\spaceskip=0pt\relax}
\providecommand{\BIBentryALTinterwordstretchfactor}{4}
\providecommand{\BIBentryALTinterwordspacing}{\spaceskip=\fontdimen2\font plus
\BIBentryALTinterwordstretchfactor\fontdimen3\font minus
  \fontdimen4\font\relax}
\providecommand{\BIBforeignlanguage}[2]{{%
\expandafter\ifx\csname l@#1\endcsname\relax
\typeout{** WARNING: IEEEtran.bst: No hyphenation pattern has been}%
\typeout{** loaded for the language `#1'. Using the pattern for}%
\typeout{** the default language instead.}%
\else
\language=\csname l@#1\endcsname
\fi
#2}}
\providecommand{\BIBdecl}{\relax}
\BIBdecl

\bibitem{debug}
\BIBentryALTinterwordspacing
T.~D. LaToza, G.~Venolia, and R.~DeLine, ``Maintaining mental models: a study
  of developer work habits,'' in \emph{28th International Conference on
  Software Engineering {(ICSE} 2006), Shanghai, China, May 20-28, 2006}, L.~J.
  Osterweil, H.~D. Rombach, and M.~L. Soffa, Eds.\hskip 1em plus 0.5em minus
  0.4em\relax {ACM}, 2006, pp. 492--501. [Online]. Available:
  \url{https://doi.org/10.1145/1134285.1134355}
\BIBentrySTDinterwordspacing

\bibitem{DBLP:journals/cacm/GouesPR19}
\BIBentryALTinterwordspacing
C.~L. Goues, M.~Pradel, and A.~Roychoudhury, ``Automated program repair,''
  \emph{Commun. {ACM}}, vol.~62, no.~12, pp. 56--65, 2019. [Online]. Available:
  \url{https://doi.org/10.1145/3318162}
\BIBentrySTDinterwordspacing

\bibitem{DBLP:journals/csur/Monperrus18}
\BIBentryALTinterwordspacing
M.~Monperrus, ``Automatic software repair: {A} bibliography,'' \emph{{ACM}
  Comput. Surv.}, vol.~51, no.~1, pp. 17:1--17:24, 2018. [Online]. Available:
  \url{https://doi.org/10.1145/3105906}
\BIBentrySTDinterwordspacing

\bibitem{livingapr}
\BIBentryALTinterwordspacing
------, ``The living review on automated program repair,'' Dec. 2020, working
  paper or preprint. [Online]. Available:
  \url{https://hal.archives-ouvertes.fr/hal-01956501}
\BIBentrySTDinterwordspacing

\bibitem{rewardrepair}
\BIBentryALTinterwordspacing
H.~Ye, M.~Martinez, and M.~Monperrus, ``Neural program repair with
  execution-based backpropagation,'' in \emph{Proceedings of the International
  Conference on Software Engineering}, 2022. [Online]. Available:
  \url{http://arxiv.org/pdf/2105.04123}
\BIBentrySTDinterwordspacing

\bibitem{chen2018sequencer}
Z.~Chen, S.~Kommrusch, M.~Tufano, L.-N. Pouchet, D.~Poshyvanyk, and
  M.~Monperrus, ``{SequenceR: Sequence-to-Sequence Learning for End-to-End
  Program Repair},'' \emph{TSE}, 2019.

\bibitem{lutellier2020coconut}
\BIBentryALTinterwordspacing
T.~Lutellier, H.~V. Pham, L.~Pang, Y.~Li, M.~Wei, and L.~Tan, ``Coconut:
  Combining context-aware neural translation models using ensemble for program
  repair,'' in \emph{Proceedings of the 29th ACM SIGSOFT International
  Symposium on Software Testing and Analysis}, ser. ISSTA 2020.\hskip 1em plus
  0.5em minus 0.4em\relax New York, NY, USA: Association for Computing
  Machinery, 2020, p. 101–114. [Online]. Available:
  \url{https://doi.org/10.1145/3395363.3397369}
\BIBentrySTDinterwordspacing

\bibitem{jiang2021cure}
N.~Jiang, T.~Lutellier, and L.~Tan, ``Cure: Code-aware neural machine
  translation for automatic program repair,'' in \emph{2021 IEEE/ACM 43rd
  International Conference on Software Engineering (ICSE)}, 2021, pp.
  1161--1173.

\bibitem{hopity}
\BIBentryALTinterwordspacing
E.~Dinella, H.~Dai, Z.~Li, M.~Naik, L.~Song, and K.~Wang, ``Hoppity: Learning
  graph transformations to detect and fix bugs in programs,'' in \emph{8th
  International Conference on Learning Representations, {ICLR} 2020, Addis
  Ababa, Ethiopia, April 26-30, 2020}.\hskip 1em plus 0.5em minus 0.4em\relax
  OpenReview.net, 2020. [Online]. Available:
  \url{https://openreview.net/forum?id=SJeqs6EFvB}
\BIBentrySTDinterwordspacing

\bibitem{zhu2021recoder}
\BIBentryALTinterwordspacing
Q.~Zhu, Z.~Sun, Y.-a. Xiao, W.~Zhang, K.~Yuan, Y.~Xiong, and L.~Zhang, ``A
  syntax-guided edit decoder for neural program repair,'' in \emph{Proceedings
  of the 29th ACM Joint Meeting on European Software Engineering Conference and
  Symposium on the Foundations of Software Engineering}.\hskip 1em plus 0.5em
  minus 0.4em\relax New York, NY, USA: Association for Computing Machinery,
  2021, p. 341–353. [Online]. Available:
  \url{https://doi.org/10.1145/3468264.3468544}
\BIBentrySTDinterwordspacing

\bibitem{transfer-vul}
\BIBentryALTinterwordspacing
Z.~Chen, S.~Kommrusch, and M.~Monperrus, ``Neural transfer learning for
  repairing security vulnerabilities in c code,'' \emph{IEEE Transactions on
  Software Engineering}, 2022. [Online]. Available:
  \url{http://arxiv.org/pdf/2104.08308}
\BIBentrySTDinterwordspacing

\bibitem{selfapr}
\BIBentryALTinterwordspacing
H.~Ye, M.~Martinez, X.~Luo, T.~Zhang, and M.~Monperrus, ``Selfapr:
  Self-supervised program repair with test execution diagnostics,'' in
  \emph{Proceedings of ASE}, 2022. [Online]. Available:
  \url{http://arxiv.org/pdf/2203.12755}
\BIBentrySTDinterwordspacing

\bibitem{defects4j}
R.~Just, D.~Jalali, and M.~D. Ernst, ``{Defects4J: A Database of Existing
  Faults to Enable Controlled Testing Studies for Java Programs},'' in
  \emph{ISSTA}, 2014, pp. 437--440.

\bibitem{quixbugs}
D.~Lin, J.~Koppel, A.~Chen, and A.~Solar-Lezama, ``{QuixBugs: A Multi-Lingual
  Program Repair Benchmark Set Based on the Quixey Challenge},'' in
  \emph{SPLASH}, 2017, p. 55–56.

\bibitem{firstlogic}
\BIBentryALTinterwordspacing
Z.~Hu, X.~Ma, Z.~Liu, E.~Hovy, and E.~Xing, ``Harnessing deep neural networks
  with logic rules,'' in \emph{Proceedings of the 54th Annual Meeting of the
  Association for Computational Linguistics (Volume 1: Long Papers)}.\hskip 1em
  plus 0.5em minus 0.4em\relax Berlin, Germany: Association for Computational
  Linguistics, Aug. 2016, pp. 2410--2420. [Online]. Available:
  \url{https://aclanthology.org/P16-1228}
\BIBentrySTDinterwordspacing

\bibitem{tufano2018src2abs}
M.~Tufano, C.~Watson, G.~Bavota, M.~D. Penta, M.~White, and D.~Poshyvanyk, ``An
  empirical study on learning bug-fixing patches in the wild via neural machine
  translation,'' \emph{CoRR}, vol. abs/1812.08693, 2018.

\bibitem{tufano2019src2abs}
M.~Tufano, J.~Pantiuchina, C.~Watson, G.~Bavota, and D.~Poshyvanyk, ``On
  learning meaningful code changes via neural machine translation,'' in
  \emph{Proceedings of the 41st International Conference on Software
  Engineering}, ser. ICSE '19, 2019.

\bibitem{Yun2019graphtransformer}
\BIBentryALTinterwordspacing
S.~Yun, M.~Jeong, R.~Kim, J.~Kang, and H.~J. Kim, ``Graph transformer
  networks,'' vol. abs/1911.06455, 2019. [Online]. Available:
  \url{http://arxiv.org/abs/1911.06455}
\BIBentrySTDinterwordspacing

\bibitem{Vijay2020graph-transfrmer}
\BIBentryALTinterwordspacing
V.~P. Dwivedi and X.~Bresson, ``A generalization of transformer networks to
  graphs,'' \emph{CoRR}, vol. abs/2012.09699, 2020. [Online]. Available:
  \url{https://arxiv.org/abs/2012.09699}
\BIBentrySTDinterwordspacing

\bibitem{treedec-markup}
\BIBentryALTinterwordspacing
J.~Zhang, J.~Du, Y.~Yang, Y.-Z. Song, S.~Wei, and L.~Dai, ``A tree-structured
  decoder for image-to-markup generation,'' in \emph{Proceedings of the 37th
  International Conference on Machine Learning}, ser. Proceedings of Machine
  Learning Research, H.~D. III and A.~Singh, Eds., vol. 119.\hskip 1em plus
  0.5em minus 0.4em\relax PMLR, 13--18 Jul 2020, pp. 11\,076--11\,085.
  [Online]. Available: \url{https://proceedings.mlr.press/v119/zhang20g.html}
\BIBentrySTDinterwordspacing

\bibitem{transformer}
\BIBentryALTinterwordspacing
A.~Vaswani, N.~Shazeer, N.~Parmar, J.~Uszkoreit, L.~Jones, A.~N. Gomez,
  L.~Kaiser, and I.~Polosukhin, ``Attention is all you need,'' vol.
  abs/1706.03762, 2017. [Online]. Available:
  \url{http://arxiv.org/abs/1706.03762}
\BIBentrySTDinterwordspacing

\bibitem{position}
\BIBentryALTinterwordspacing
J.~Gehring, M.~Auli, D.~Grangier, D.~Yarats, and Y.~N. Dauphin, ``Convolutional
  sequence to sequence learning,'' \emph{CoRR}, vol. abs/1705.03122, 2017.
  [Online]. Available: \url{http://arxiv.org/abs/1705.03122}
\BIBentrySTDinterwordspacing

\bibitem{layernorm}
\BIBentryALTinterwordspacing
L.~J. Ba, J.~R. Kiros, and G.~E. Hinton, ``Layer normalization,'' \emph{CoRR},
  vol. abs/1607.06450, 2016. [Online]. Available:
  \url{http://arxiv.org/abs/1607.06450}
\BIBentrySTDinterwordspacing

\bibitem{treedec-nmt}
\BIBentryALTinterwordspacing
X.~Wang, H.~Pham, P.~Yin, and G.~Neubig, ``A tree-based decoder for neural
  machine translation,'' \emph{CoRR}, vol. abs/1808.09374, 2018. [Online].
  Available: \url{http://arxiv.org/abs/1808.09374}
\BIBentrySTDinterwordspacing

\bibitem{pointer}
\BIBentryALTinterwordspacing
O.~Vinyals, M.~Fortunato, and N.~Jaitly, ``Pointer networks,'' 2015. [Online].
  Available: \url{https://arxiv.org/abs/1506.03134}
\BIBentrySTDinterwordspacing

\bibitem{crossentropy}
\BIBentryALTinterwordspacing
G.~Cybenko, D.~O'Leary, and J.~Rissanen, \emph{The Mathematics of Information
  Coding, Extraction and Distribution}, ser. The IMA Volumes in Mathematics and
  its Applications.\hskip 1em plus 0.5em minus 0.4em\relax Springer New York,
  1998. [Online]. Available:
  \url{https://books.google.com/books?id=jDrp4QEGioMC}
\BIBentrySTDinterwordspacing

\bibitem{kldivergence}
\BIBentryALTinterwordspacing
S.~Kullback and R.~A. Leibler, ``{On Information and Sufficiency},'' \emph{The
  Annals of Mathematical Statistics}, vol.~22, no.~1, pp. 79 -- 86, 1951.
  [Online]. Available: \url{https://doi.org/10.1214/aoms/1177729694}
\BIBentrySTDinterwordspacing

\bibitem{ensemble1}
R.~Polikar, ``Ensemble based systems in decision making,'' \emph{IEEE Circuits
  and Systems Magazine}, vol.~6, no.~3, pp. 21--45, 2006.

\bibitem{ensemble2}
L.~Rokach, ``Ensemble-based classifiers,'' \emph{Artificial Intelligence
  Review}, vol.~33, pp. 1--39, 2009.

\bibitem{yang2017validate}
\BIBentryALTinterwordspacing
J.~Yang, A.~Zhikhartsev, Y.~Liu, and L.~Tan, ``Better test cases for better
  automated program repair,'' in \emph{FSE}, ser. ESEC/FSE 2017.\hskip 1em plus
  0.5em minus 0.4em\relax ACM, 2017, p. 831–841. [Online]. Available:
  \url{https://doi.org/10.1145/3106237.3106274}
\BIBentrySTDinterwordspacing

\bibitem{simfix}
\BIBentryALTinterwordspacing
J.~Jiang, Y.~Xiong, H.~Zhang, Q.~Gao, and X.~Chen, ``Shaping program repair
  space with existing patches and similar code,'' in \emph{Proceedings of the
  27th ACM SIGSOFT International Symposium on Software Testing and Analysis},
  ser. ISSTA 2018.\hskip 1em plus 0.5em minus 0.4em\relax New York, NY, USA:
  Association for Computing Machinery, 2018, p. 298–309. [Online]. Available:
  \url{https://doi.org/10.1145/3213846.3213871}
\BIBentrySTDinterwordspacing

\bibitem{tbar}
K.~Liu, A.~Koyuncu, D.~Kim, and T.~F. Bissyand{\'e}, ``{TBar: Revisiting
  Template-Based Automated Program Repair},'' in \emph{ISSTA}.\hskip 1em plus
  0.5em minus 0.4em\relax ACM, 2019.

\bibitem{efficiency}
\BIBentryALTinterwordspacing
K.~Liu, S.~Wang, A.~Koyuncu, K.~Kim, T.~F. Bissyand{\'{e}}, D.~Kim, P.~Wu,
  J.~Klein, X.~Mao, and Y.~L. Traon, ``On the efficiency of test suite based
  program repair: {A} systematic assessment of 16 automated repair systems for
  java programs,'' \emph{CoRR}, vol. abs/2008.00914, 2020. [Online]. Available:
  \url{https://arxiv.org/abs/2008.00914}
\BIBentrySTDinterwordspacing

\bibitem{Li2020dlfix}
Y.~Li, S.~Wang, and T.~N. Nguyen, ``{DLFix: Context-Based Code Transformation
  Learning for Automated Program Repair},'' in \emph{ICSE}.\hskip 1em plus
  0.5em minus 0.4em\relax ACM, 2020, p. 602–614.

\bibitem{ochiai}
R.~Abreu, P.~Zoeteweij, and A.~J. van Gemund, ``On the accuracy of
  spectrum-based fault localization,'' in \emph{Testing: Academic and
  Industrial Conference Practice and Research Techniques - MUTATION
  (TAICPART-MUTATION 2007)}, 2007, pp. 89--98.

\bibitem{2019javalang}
\BIBentryALTinterwordspacing
C.~Thunes, ``javalang,'' 2020. [Online]. Available:
  \url{https://github.com/c2nes/javalang}
\BIBentrySTDinterwordspacing

\bibitem{smith2019javaparser}
\BIBentryALTinterwordspacing
N.~Smith, D.~Van~Bruggen, and F.~Tomassetti, ``Javaparser: Visited,'' 2019.
  [Online]. Available: \url{https://github.com/javaparser/javaparser}
\BIBentrySTDinterwordspacing

\bibitem{pytorch}
\BIBentryALTinterwordspacing
``Pytorch,'' 2022. [Online]. Available: \url{https://pytorch.org/}
\BIBentrySTDinterwordspacing

\bibitem{beamsearch}
\BIBentryALTinterwordspacing
I.~Sutskever, O.~Vinyals, and Q.~V. Le, ``Sequence to sequence learning with
  neural networks,'' \emph{CoRR}, vol. abs/1409.3215, 2014. [Online].
  Available: \url{http://arxiv.org/abs/1409.3215}
\BIBentrySTDinterwordspacing

\bibitem{hercules}
S.~{Saha}, R.~k.~{Saha}, and M.~r.~{Prasad}, ``{Harnessing Evolution for
  Multi-Hunk Program Repair},'' in \emph{ICSE}.\hskip 1em plus 0.5em minus
  0.4em\relax IEEE, 2019, pp. 13--24.

\bibitem{DBLP:journals/tse/YuanB20/arja}
\BIBentryALTinterwordspacing
Y.~Yuan and W.~Banzhaf, ``{ARJA:} automated repair of java programs via
  multi-objective genetic programming,'' \emph{{IEEE} Trans. Software Eng.},
  vol.~46, no.~10, pp. 1040--1067, 2020. [Online]. Available:
  \url{https://doi.org/10.1109/TSE.2018.2874648}
\BIBentrySTDinterwordspacing

\bibitem{DBLP:conf/icse/WenCWHC18}
\BIBentryALTinterwordspacing
M.~Wen, J.~Chen, R.~Wu, D.~Hao, and S.~Cheung, ``Context-aware patch generation
  for better automated program repair,'' in \emph{Proceedings of the 40th
  International Conference on Software Engineering, {ICSE} 2018, Gothenburg,
  Sweden, May 27 - June 03, 2018}, M.~Chaudron, I.~Crnkovic, M.~Chechik, and
  M.~Harman, Eds.\hskip 1em plus 0.5em minus 0.4em\relax {ACM}, 2018, pp.
  1--11. [Online]. Available: \url{https://doi.org/10.1145/3180155.3180233}
\BIBentrySTDinterwordspacing

\bibitem{DBLP:conf/kbse/SahaLYP17/elixir}
\BIBentryALTinterwordspacing
R.~K. Saha, Y.~Lyu, H.~Yoshida, and M.~R. Prasad, ``{ELIXIR:} effective object
  oriented program repair,'' in \emph{Proceedings of the 32nd {IEEE/ACM}
  International Conference on Automated Software Engineering, {ASE} 2017,
  Urbana, IL, USA, October 30 - November 03, 2017}, G.~Rosu, M.~D. Penta, and
  T.~N. Nguyen, Eds.\hskip 1em plus 0.5em minus 0.4em\relax {IEEE} Computer
  Society, 2017, pp. 648--659. [Online]. Available:
  \url{https://doi.org/10.1109/ASE.2017.8115675}
\BIBentrySTDinterwordspacing

\bibitem{DBLP:conf/issta/GhanbariBZ19}
\BIBentryALTinterwordspacing
A.~Ghanbari, S.~Benton, and L.~Zhang, ``Practical program repair via bytecode
  mutation,'' in \emph{Proceedings of the 28th {ACM} {SIGSOFT} International
  Symposium on Software Testing and Analysis, {ISSTA} 2019, Beijing, China,
  July 15-19, 2019}, D.~Zhang and A.~M{\o}ller, Eds.\hskip 1em plus 0.5em minus
  0.4em\relax {ACM}, 2019, pp. 19--30. [Online]. Available:
  \url{https://doi.org/10.1145/3293882.3330559}
\BIBentrySTDinterwordspacing

\bibitem{DBLP:conf/icse/XiongWYZH0017/acs}
\BIBentryALTinterwordspacing
Y.~Xiong, J.~Wang, R.~Yan, J.~Zhang, S.~Han, G.~Huang, and L.~Zhang, ``Precise
  condition synthesis for program repair,'' in \emph{Proceedings of the 39th
  International Conference on Software Engineering, {ICSE} 2017, Buenos Aires,
  Argentina, May 20-28, 2017}, S.~Uchitel, A.~Orso, and M.~P. Robillard,
  Eds.\hskip 1em plus 0.5em minus 0.4em\relax {IEEE} / {ACM}, 2017, pp.
  416--426. [Online]. Available: \url{https://doi.org/10.1109/ICSE.2017.45}
\BIBentrySTDinterwordspacing

\bibitem{DBLP:journals/tse/XuanMDCMDBM17/nopol}
\BIBentryALTinterwordspacing
J.~Xuan, M.~Martinez, F.~Demarco, M.~Clement, S.~R.~L. Marcote, T.~Durieux,
  D.~L. Berre, and M.~Monperrus, ``Nopol: Automatic repair of conditional
  statement bugs in java programs,'' \emph{{IEEE} Trans. Software Eng.},
  vol.~43, no.~1, pp. 34--55, 2017. [Online]. Available:
  \url{https://doi.org/10.1109/TSE.2016.2560811}
\BIBentrySTDinterwordspacing

\bibitem{DBLP:conf/ssbse/MartinezM18/cardumen}
\BIBentryALTinterwordspacing
M.~Martinez and M.~Monperrus, ``Ultra-large repair search space with
  automatically mined templates: The cardumen mode of astor,'' in
  \emph{Search-Based Software Engineering - 10th International Symposium,
  {SSBSE} 2018, Montpellier, France, September 8-9, 2018, Proceedings}, ser.
  Lecture Notes in Computer Science, T.~E. Colanzi and P.~McMinn, Eds., vol.
  11036.\hskip 1em plus 0.5em minus 0.4em\relax Springer, 2018, pp. 65--86.
  [Online]. Available: \url{https://doi.org/10.1007/978-3-319-99241-9\_3}
\BIBentrySTDinterwordspacing

\bibitem{deepdebug}
\BIBentryALTinterwordspacing
D.~Drain, C.~B. Clement, G.~Serrato, and N.~Sundaresan, ``Deepdebug: Fixing
  python bugs using stack traces, backtranslation, and code skeletons,''
  \emph{CoRR}, vol. abs/2105.09352, 2021. [Online]. Available:
  \url{https://arxiv.org/abs/2105.09352}
\BIBentrySTDinterwordspacing

\bibitem{DBLP:conf/icse/TarlowMRCMSA20}
\BIBentryALTinterwordspacing
D.~Tarlow, S.~Moitra, A.~Rice, Z.~Chen, P.~Manzagol, C.~Sutton, and
  E.~Aftandilian, ``Learning to fix build errors with graph2diff neural
  networks,'' in \emph{{ICSE} '20: 42nd International Conference on Software
  Engineering, Workshops, Seoul, Republic of Korea, 27 June - 19 July,
  2020}.\hskip 1em plus 0.5em minus 0.4em\relax {ACM}, 2020, pp. 19--20.
  [Online]. Available: \url{https://doi.org/10.1145/3387940.3392181}
\BIBentrySTDinterwordspacing

\bibitem{codit}
S.~Chakraborty, Y.~Ding, M.~Allamanis, and B.~Ray, ``Codit: Code editing with
  tree-based neural models,'' \emph{IEEE Transactions on Software Engineering},
  vol.~48, no.~4, pp. 1385--1399, 2022.

\bibitem{DBLP:conf/acl/LingBGHKWS16}
\BIBentryALTinterwordspacing
W.~Ling, P.~Blunsom, E.~Grefenstette, K.~M. Hermann, T.~Kocisk{\'{y}}, F.~Wang,
  and A.~W. Senior, ``Latent predictor networks for code generation,'' in
  \emph{Proceedings of the 54th Annual Meeting of the Association for
  Computational Linguistics, {ACL} 2016, August 7-12, 2016, Berlin, Germany,
  Volume 1: Long Papers}.\hskip 1em plus 0.5em minus 0.4em\relax The
  Association for Computer Linguistics, 2016. [Online]. Available:
  \url{https://doi.org/10.18653/v1/p16-1057}
\BIBentrySTDinterwordspacing

\bibitem{DBLP:conf/acl/DongL16}
\BIBentryALTinterwordspacing
L.~Dong and M.~Lapata, ``Language to logical form with neural attention,'' in
  \emph{Proceedings of the 54th Annual Meeting of the Association for
  Computational Linguistics, {ACL} 2016, August 7-12, 2016, Berlin, Germany,
  Volume 1: Long Papers}.\hskip 1em plus 0.5em minus 0.4em\relax The
  Association for Computer Linguistics, 2016. [Online]. Available:
  \url{https://doi.org/10.18653/v1/p16-1004}
\BIBentrySTDinterwordspacing

\bibitem{DBLP:conf/acl/YinN17}
\BIBentryALTinterwordspacing
P.~Yin and G.~Neubig, ``A syntactic neural model for general-purpose code
  generation,'' in \emph{Proceedings of the 55th Annual Meeting of the
  Association for Computational Linguistics, {ACL} 2017, Vancouver, Canada,
  July 30 - August 4, Volume 1: Long Papers}, R.~Barzilay and M.~Kan,
  Eds.\hskip 1em plus 0.5em minus 0.4em\relax Association for Computational
  Linguistics, 2017, pp. 440--450. [Online]. Available:
  \url{https://doi.org/10.18653/v1/P17-1041}
\BIBentrySTDinterwordspacing

\bibitem{DBLP:conf/emnlp/YinN18}
\BIBentryALTinterwordspacing
------, ``{TRANX:} {A} transition-based neural abstract syntax parser for
  semantic parsing and code generation,'' in \emph{Proceedings of the 2018
  Conference on Empirical Methods in Natural Language Processing, {EMNLP} 2018:
  System Demonstrations, Brussels, Belgium, October 31 - November 4, 2018},
  E.~Blanco and W.~Lu, Eds.\hskip 1em plus 0.5em minus 0.4em\relax Association
  for Computational Linguistics, 2018, pp. 7--12. [Online]. Available:
  \url{https://doi.org/10.18653/v1/d18-2002}
\BIBentrySTDinterwordspacing

\bibitem{DBLP:conf/acl/RabinovichSK17}
\BIBentryALTinterwordspacing
M.~Rabinovich, M.~Stern, and D.~Klein, ``Abstract syntax networks for code
  generation and semantic parsing,'' in \emph{Proceedings of the 55th Annual
  Meeting of the Association for Computational Linguistics, {ACL} 2017,
  Vancouver, Canada, July 30 - August 4, Volume 1: Long Papers}, R.~Barzilay
  and M.~Kan, Eds.\hskip 1em plus 0.5em minus 0.4em\relax Association for
  Computational Linguistics, 2017, pp. 1139--1149. [Online]. Available:
  \url{https://doi.org/10.18653/v1/P17-1105}
\BIBentrySTDinterwordspacing

\bibitem{DBLP:conf/aaai/SunZXSMZ20}
\BIBentryALTinterwordspacing
Z.~Sun, Q.~Zhu, Y.~Xiong, Y.~Sun, L.~Mou, and L.~Zhang, ``Treegen: {A}
  tree-based transformer architecture for code generation,'' in \emph{The
  Thirty-Fourth {AAAI} Conference on Artificial Intelligence, {AAAI} 2020, The
  Thirty-Second Innovative Applications of Artificial Intelligence Conference,
  {IAAI} 2020, The Tenth {AAAI} Symposium on Educational Advances in Artificial
  Intelligence, {EAAI} 2020, New York, NY, USA, February 7-12, 2020}.\hskip 1em
  plus 0.5em minus 0.4em\relax {AAAI} Press, 2020, pp. 8984--8991. [Online].
  Available: \url{https://ojs.aaai.org/index.php/AAAI/article/view/6430}
\BIBentrySTDinterwordspacing

\bibitem{DBLP:conf/iclr/YaoXYSN21}
\BIBentryALTinterwordspacing
Z.~Yao, F.~F. Xu, P.~Yin, H.~Sun, and G.~Neubig, ``Learning structural edits
  via incremental tree transformations,'' in \emph{9th International Conference
  on Learning Representations, {ICLR} 2021, Virtual Event, Austria, May 3-7,
  2021}.\hskip 1em plus 0.5em minus 0.4em\relax OpenReview.net, 2021. [Online].
  Available: \url{https://openreview.net/forum?id=v9hAX77--cZ}
\BIBentrySTDinterwordspacing

\bibitem{DBLP:conf/iclr/YinNABG19}
\BIBentryALTinterwordspacing
P.~Yin, G.~Neubig, M.~Allamanis, M.~Brockschmidt, and A.~L. Gaunt, ``Learning
  to represent edits,'' in \emph{7th International Conference on Learning
  Representations, {ICLR} 2019, New Orleans, LA, USA, May 6-9, 2019}.\hskip 1em
  plus 0.5em minus 0.4em\relax OpenReview.net, 2019. [Online]. Available:
  \url{https://openreview.net/forum?id=BJl6AjC5F7}
\BIBentrySTDinterwordspacing

\bibitem{DBLP:journals/pacmpl/Brody0Y20}
\BIBentryALTinterwordspacing
S.~Brody, U.~Alon, and E.~Yahav, ``A structural model for contextual code
  changes,'' \emph{Proc. {ACM} Program. Lang.}, vol.~4, no. {OOPSLA}, pp.
  215:1--215:28, 2020. [Online]. Available:
  \url{https://doi.org/10.1145/3428283}
\BIBentrySTDinterwordspacing

\bibitem{codebert}
\BIBentryALTinterwordspacing
Z.~Feng, D.~Guo, D.~Tang, N.~Duan, X.~Feng, M.~Gong, L.~Shou, B.~Qin, T.~Liu,
  D.~Jiang, and M.~Zhou, ``Codebert: {A} pre-trained model for programming and
  natural languages,'' \emph{CoRR}, vol. abs/2002.08155, 2020. [Online].
  Available: \url{https://arxiv.org/abs/2002.08155}
\BIBentrySTDinterwordspacing

\bibitem{codet5}
W.~Yue, W.~Weishi, J.~Shafiq, and C.~H. Steven, ``Codet5: Identifier-aware
  unified pre-trained encoder-decoder models for code understanding and
  generation,'' in \emph{Proceedings of the 2021 Conference on Empirical
  Methods in Natural Language Processing, EMNLP 2021}, 2021.

\bibitem{codegen}
E.~Nijkamp, B.~Pang, H.~Hayashi, L.~Tu, H.~Wang, Y.~Zhou, S.~Savarese, and
  C.~Xiong, ``A conversational paradigm for program synthesis,'' \emph{arXiv
  preprint}, 2022.

\bibitem{incoder}
\BIBentryALTinterwordspacing
D.~Fried, A.~Aghajanyan, J.~Lin, S.~Wang, E.~Wallace, F.~Shi, R.~Zhong, W.-t.
  Yih, L.~Zettlemoyer, and M.~Lewis, ``Incoder: A generative model for code
  infilling and synthesis,'' 2022. [Online]. Available:
  \url{https://arxiv.org/abs/2204.05999}
\BIBentrySTDinterwordspacing

\bibitem{codex}
\BIBentryALTinterwordspacing
M.~Chen, J.~Tworek, H.~Jun, Q.~Yuan, H.~P. de~Oliveira~Pinto, J.~Kaplan,
  H.~Edwards, Y.~Burda, N.~Joseph, G.~Brockman, A.~Ray, R.~Puri, G.~Krueger,
  M.~Petrov, H.~Khlaaf, G.~Sastry, P.~Mishkin, B.~Chan, S.~Gray, N.~Ryder,
  M.~Pavlov, A.~Power, L.~Kaiser, M.~Bavarian, C.~Winter, P.~Tillet, F.~P.
  Such, D.~Cummings, M.~Plappert, F.~Chantzis, E.~Barnes, A.~Herbert{-}Voss,
  W.~H. Guss, A.~Nichol, A.~Paino, N.~Tezak, J.~Tang, I.~Babuschkin, S.~Balaji,
  S.~Jain, W.~Saunders, C.~Hesse, A.~N. Carr, J.~Leike, J.~Achiam, V.~Misra,
  E.~Morikawa, A.~Radford, M.~Knight, M.~Brundage, M.~Murati, K.~Mayer,
  P.~Welinder, B.~McGrew, D.~Amodei, S.~McCandlish, I.~Sutskever, and
  W.~Zaremba, ``Evaluating large language models trained on code,''
  \emph{CoRR}, vol. abs/2107.03374, 2021. [Online]. Available:
  \url{https://arxiv.org/abs/2107.03374}
\BIBentrySTDinterwordspacing

\bibitem{share}
\BIBentryALTinterwordspacing
``Replication package of this work,'' 2022. [Online]. Available:
  \url{https://github.com/lin-tan/knod}
\BIBentrySTDinterwordspacing

\end{thebibliography}

\end{document}